\documentclass{article}

\usepackage{arxiv}

\usepackage[utf8]{inputenc} 
\usepackage[T1]{fontenc}    
\usepackage{hyperref}       
\usepackage{url}            
\usepackage{booktabs}       
\usepackage{amsfonts}       
\usepackage{nicefrac}       
\usepackage{microtype}      
\usepackage{lipsum}		
\usepackage{graphicx}
\usepackage{natbib}
\usepackage{doi}
\usepackage{longtable}
\usepackage{booktabs}
\usepackage{array}
\usepackage{ragged2e}
\usepackage{etoolbox}
\usepackage{textcomp}
\usepackage{amsmath}
\usepackage{algorithm}
\usepackage{algorithmic}
\usepackage{amssymb, bbm}

\title{pvEBayes: An R Package for Empirical Bayes Methods in Pharmacovigilance}

\author{
 Yihao Tan \\
  Department of Biostatistics\\
  School of Public Health and Health Professions\\
  State University of New York at Buffalo\\
  Buffalo, New York, USA\\
   \And
 Marianthi Markatou \\
  Department of Biostatistics\\
  School of Public Health and Health Professions\\
  State University of New York at Buffalo\\
  Buffalo, New York, USA\\
  \texttt{markatou@buffalo.edu} \\
  \And
 Saptarshi Chakraborty\thanks{corresponding author} \\
  Department of Biostatistics\\
  School of Public Health and Health Professions\\
  State University of New York at Buffalo\\
  Buffalo, New York, USA\\
  \texttt{chakrab2@buffalo.edu} \\
}

\renewcommand{\shorttitle}{}

\shortauthors{Tan, Markatou, and Chakraborty}
\begin{document}

\maketitle

\begin{abstract}%
Monitoring the safety of medical products is a core concern of contemporary pharmacovigilance. To support drug safety assessment, Spontaneous Reporting Systems (SRS) collect reports of suspected adverse events of approved medical products offering a critical resource for identifying potential safety concerns that may not emerge during clinical trials. Modern nonparametric empirical Bayes methods are flexible statistical approaches that can accurately identify and estimate the strength of the association between an adverse event and a drug from SRS data. However, there is currently no comprehensive and easily accessible implementation of these methods. Here, we introduce the R package pvEBayes, which implements a suite of nonparametric empirical Bayes methods for pharmacovigilance, along with post-processing tools and graphical summaries for streamlining the application of these methods. Detailed examples are provided to demonstrate the application of the package through analyses of two real-world SRS datasets curated from the publicly available FDA FAERS database.
\end{abstract}


\section{Introduction}\label{introduction}

Monitoring the safety of medical products after they enter the market is a fundamental aim of pharmacovigilance. It involves detecting and analyzing adverse events (AEs) associated with drugs and medical treatments\citep{amery1999there}. While clinical trials provide initial safety evaluations, the identification of serious, rare adverse events is often limited by factors such as sample size, study duration, and scope. As a result, some AE-drug interactions may only become evident in larger, more diverse populations after regulatory approval. To address this, Spontaneous Reporting Systems (SRS) have been established worldwide, collecting reports of suspected drug-related AEs from healthcare professionals, pharmaceutical companies, and patients, and serving as a key resource for post-market drug safety assessment. One of the largest and most widely used SRS databases is the FDA's Adverse Event Reporting System (FAERS).

Despite their usefulness, SRS data are inherently observational and present several challenges, including underreporting of AEs, lack of proper controls, inaccuracies in measuring drug use, and the presence of selection bias and confounding\citep{markatou2014pattern}. Due to these limitations, direct causal inference between drugs and AEs is difficult. Instead, pharmacovigilance typically relies on SRS data mining methods to analyze AE-drug associations. These methods commonly evaluate whether the observed frequency (\(O\)) of an AE-drug pair significantly exceeds its expected frequency (\(E\)), where \(E\) represents the baseline count assuming no association of the corresponding AE-drug pair exists. If the ratio \(O/E\) is substantially greater than one, the AE-drug combination is identified as a signal, warranting further investigation.

Over the past decades, a variety of statistical methods have been developed to detect potential signals in SRS data, including proportional reporting ratios (PRR)\citep{evans2001use}, reporting odds ratio (ROR) \citep{rothman2004reporting}, likelihood ratio test (LRT) based methods \citep{ding2020evaluation, huang2011likelihood, chakraborty2022use, zhao2018extended, huang2017zero}, and Bayesian/empirical Bayesian methods \citep{huang2013likelihood, bate1998bayesian, dumouchel1999bayesian, hu2015signal, tan2025flexibleempiricalbayesianapproaches}. PRR and ROR are heuristic methods that use predefined thresholds on \(\{O/E\}\) values to identify potential AE-drug signals. More formal methods parametrize the \(\{O/E\}\), or some functions of \(\{O/E\}\) values, with probabilistic models for the observed report counts. These methods provide more principled signal determination, using hypothesis tests with controlled type I errors and false discovery rates in frequentist settings or with prespecified posterior probabilities of being a signal from an appropriately established Bayesian model for the data. Other methods of signal detection for SRS data include the pattern discover approach MDDC\citep{liu2024mddcrpythonpackage}, sequential testing based approaches\citep{li2020adverseeventenrichmenttests, shih2010sequential}, and adaptive lasso based approach\citep{courtois2021new}.

The structure of SRS data is inherently complex, including ``signals'', ``non-signals'', and ``structural zeros''. Signals correspond to AE-drug pairs whose \(O/E\) ratio exceeds 1, indicating a potential association. The ratio \(O/E\) for signal AE-drug pairs usually takes a range of values. Non-signals, often representing the majority of AE-drug pairs in SRS data, have a \(O/E\) ratio equal to 1, suggesting no meaningful relationship between the corresponding drug and AE. Structural zeros arise in cases where specific AE-drug combinations are biologically or mechanistically impossible to occur, resulting in an absence of reports that are not merely due to underreporting but rather reflect a true zero occurrence. Most methods discussed above focus on identifying whether an AE-drug pair is a `signal' or `non-signal'. However, the limitations of a simplistic signal/non-signal dichotomy in statistical inference are increasingly recognized and acknowledged by contemporary science and biomedicine; for example, it can sometimes obscure important nuances in the data\citep{amrheinScientistsRiseStatistical2019, gelmanStatisticalCrisisScience2016, rothmanDisengagingStatisticalSignificance2016, wassersteinASAStatementPValues2016}. Bayesian approaches are well-suited for estimating signal strengths and quantifying uncertainty across AE-drug pairs in SRS data. By Bayes' rule, these methods incorporate a prior distribution, defined over the model parameters reflecting the structure of the \(O/E\) ratios in the SRS data, with the observed SRS data (likelihood) to obtain a posterior distribution for model parameters. The posterior distribution then enables probabilistic inference for model parameters while accounting for uncertainty by integrating prior knowledge and observed evidence. Specifically, the posterior inference of the \(O/E\) ratio yields a distribution that captures both the estimated signal strength and its associated uncertainty for each AE-drug pair. This allows us to differentiate between AE-drug combinations with similar detection outcomes---such as those with expected signal strengths of 1.5 and 4.0---even though both of them are significantly greater than 1 under the signal detection framework. Importantly, these signal strengths have
very different implications in medical and clinical contexts. For accurate estimation and inference of these signal strengths in a Bayesian framework, appropriate prior specification is crucial. In the following, we discuss the structure of SRS data, outline several criteria for the prior of Bayesian models that should be satisfied for flexible and robust inference, and review existing Bayesian approaches in pharmacovigilance.

Due to the complexity of SRS data, a capable prior distribution needs to be: (1) sufficiently ``vague'' and flexible to reflect the data (likelihood) information in the posterior and (2) accommodate multiple modes in the distribution of the parameters for \(O/E\) ratios. Most existing Bayesian methods---such as BCPNN \citep{bate1998bayesian}, GPS \citep{dumouchel1999bayesian}, HDP \citep{hu2015signal}, KM \citep{koenker2014convex}, the Efron approach \citep{efron2016empirical}, and the general-gamma model \citep{tan2025flexibleempiricalbayesianapproaches}---have been reported to perform well in signal detection. However, not all of these methods satisfy the key criteria outlined above. The performance of methods listed above varies in signal strength estimation \citep{tan2025flexibleempiricalbayesianapproaches}. The prior of Bayesian confidence propagation neural networks (BCPNN) model does not account for the joint variability among all \(O/E\) values. The single-gamma and prior gamma Poisson shrinker (GPS) model utilizes a mixture of two gamma prior distributions which may not be flexible enough to capture the complexity of SRS data. The hierarchical Dirichlet process (HDP) model provides flexible inference within a non-parametric full Bayesian framework but depends heavily on hyperparameters that are not straightforward to determine and require an MCMC approach for implementation which is computationally heavy. Non-parametric empirical Bayesian approaches, including the general-gamma model \citep{tan2025flexibleempiricalbayesianapproaches}, Koenker and Mizera's model\citep{koenker2014convex} and Efron's approach \citep{efron2016empirical, tan2025flexibleempiricalbayesianapproaches}, offer accurate and flexible alternatives where the prior complexity is adaptively determined from the SRS data, while still maintaining scalability for large datasets.

Currently, a few R packages available on CRAN offer functionalities relevant to medical product safety including the packages \pkg{PhViD} \citep{pkg_phvid}, \CRANpkg{openEBGM} \citep{canida2017}, \CRANpkg{pvLRT} \citep{pkg_pvLRT, chakraborty2023likelihood}, \CRANpkg{sglr} \citep{pkg_sglr}, \CRANpkg{Sequential} \citep{pkg_sequential}, \CRANpkg{AEenrich} \citep{pkg_aeenrich}, \CRANpkg{mds} \citep{pkg_mds}, \CRANpkg{MDDC} \citep{liu2024mddcrpythonpackage} and \CRANpkg{adapt4pv} \citep{pkg_adapt4pv}. Table 1 lists these packages and the methods implemented in each package. The HDP method is not available directly on any software package, but can be implemented using an MCMC sampler, e.g., JAGS or Stan. The KM method is available in the \CRANpkg{REBayes} \citep{pkg_KM} package; however, \CRANpkg{REBayes} is not explicitly developed for pharmacovigilance but rather provides functions for parameter estimation in a general nonparametric empirical Bayes setting. In addition, this package relies on Mosek\citep{Rmosek_manual}, a commercial solver for convex optimization problems, which may limit accessibility for users due to licensing requirements. \CRANpkg{deconvolveR} \citep{pkg_Efron} is a related package implementing Efron's approach for general nonparametric empirical Bayes modeling; But, it does not include functionality for pharmacovigilance modeling. Motivated by the lack of easily accessible and open-source computational solutions to the non-parametric Bayesian pharmacovigilance approaches, our package \CRANpkg{pvEBayes} \citep{pvEBayes_pkg} provides an implementation of GPS, \(K\)-gamma and general-gamma, KM and Efron models and functionalities for null value estimation, hyperparameter selection and result visualization, primarily intended for signal detection and estimation in pharmacovigilance.

The key contributions of \CRANpkg{pvEBayes} are provided below. First, \CRANpkg{pvEBayes} provides a comprehensive software framework for fitting GPS, \(K\)-gamma, general-gamma, KM, and Efron models. For models with gamma mixture priors (GPS, \(K\)-gamma, and general-gamma), parameter estimation is performed using an Expectation conditional maximization (ECM) algorithm. This algorithm is implemented in C++ for efficiency and is imported to R via \CRANpkg{Rcpp} \citep{pkg_rcpp} and \CRANpkg{RcppEigen} \citep{pkg_rcppeigen}. In addition, we provide a fully open-source KM model implementation that uses \CRANpkg{CVXR} for the model fitting. The Efron model implementation is adapted from the \CRANpkg{deconvolveR} package and modified to accommodate pharmacovigilance modeling. Formal statistical model comparison methods-Akaike Information Criterion (AIC) and Bayesian Information Criterion (BIC)-are implemented for the selection of the hyperparameters in the general-gamma method and the Efron method. Second, two summary plots, a heatmap, and an eyeplot from \CRANpkg{ggdist}, are provided for presenting signal detection and signal estimation results. These plots visualize the posterior probability of being a signal (signal detection) and the posterior median with 90\% equi-tailed credible intervals (signal estimation) for selected drugs and AEs. These plots are generated based on \CRANpkg{ggplot2} \citep{pkg_ggplot2}. As a result, their visual attributes (e.g., color schemes, text size, axis labels) can be easily customized using standard ggplot2 syntax, allowing users to tailor the appearance to their specific needs. Finally, our package includes several SRS data tables counting AE-drug reports for two groups of drugs---statins and Gadolinium-based Contrast Agents (GBCA; \cite{zhao2018extended})---as R datasets. The datasets collect AE-drug reports from FAERS for the quarters 2021Q1-2024Q4.

The remainder of the article is organized as follows. Section 2 provides a brief review of non-parametric empirical Bayes approaches to pharmacovigilance. Section 3 shows a simulation study on hyperparameter selection for the general gamma method. Section 4 examplifies \CRANpkg{pvEBayes} by analyzing two sets of real pharmacovigilance data. Section 5 concludes the article with a brief discussion.

\begin{longtable}[t]{l>{\raggedright\arraybackslash}p{6cm}>{\raggedright\arraybackslash}p{6cm}}
\caption{\label{tab:table-R-packages}Existing R packages on CRAN with functionalities for pharmacovigilance.}\\
\toprule
Package & Method & Aim\\
\midrule
\pkg{PhViD} & PRR, ROR, BCPNN, GPS & Signal detection for pharmacovigilance\\
\CRANpkg{openEBGM} & GPS method & Signal detection for pharmacovigilance\\
\CRANpkg{pvLRT} & (Pseudo) LRT approaches based on log-linear models & Signal detection for pharmacovigilance\\
\CRANpkg{sglr} & Sequential Generalized Likelihood Ratio decision boundaries & Signal detection for pharmacovigilance\\
\CRANpkg{Sequential} & Max SPRT statistic & Signal detection for pharmacovigilance\\
\addlinespace
\CRANpkg{AEenrich} & Modified Fisher’s exact test and Modified Kolmogorov Smirnov statistic & Signal detection for pharmacovigilance\\
\CRANpkg{mds} & SRS Data preprocessing & Offers functions for processing and organizing messy, unstructured SRS data\\
\CRANpkg{MDDC} & Modified Detecting Deviating Cells Algorithm & Signal detection for pharmacovigilance\\
\CRANpkg{adapt4pv} & adaptive lasso based pharmacovigilance signal detection methods & Signal detection for pharmacovigilance\\
\CRANpkg{REBayes} & KM & Offers functions for parameter estimation in a general nonparametric empirical Bayes setting\\
\addlinespace
\CRANpkg{pvEBayes} & GPS, K-gamma, General-gamma, KM and Efron & Both signal detection and signal estimation for pharmacovigilance\\
\bottomrule
\end{longtable}

\section{Notation and a brief review of the non-parametric empirical Bayes approaches to pharmacovigilance}\label{notation-and-a-brief-review-of-the-non-parametric-empirical-bayes-approaches-to-pharmacovigilance}

This section introduces our notation and reviews existing non-parametric empirical Bayes approaches to pharmacovigilance. We consider an SRS dataset cataloging AE reports on \(I\) rows across \(J\) columns, where drugs are reported. Let \(N_{ij}\) denote the number of reported cases for the \(i\)-th AE and the \(j\)-th drug, where \(i = 1, \dots, I\) and \(j = 1, \dots, J\). We consider the \(J\)-th column as the reference category, often labeled as ``Other drugs,'' against which the prevalence of AEs in the remaining drugs (\(j = 1, \dots, J-1\)) is assessed. Similarly, the \(I\)-th row is the reference category, referred to as ``Other AEs.'' These reference drug and AE categories commonly appear in SRS databases as natural comparators; if absent, they can be constructed by collapsing or grouping some existing AEs or drugs that are not in the same class with drugs or AEs under investigation. These AE-drug pairwise occurrences are summarized into an \(I \times J\) contingency table, where the \((i, j)\)-th cell catalogs the observed count \(N_{ij}\) indicating the number of cases involving \(i\)-th AE and the \(j\)-th drug. The row, column (marginal), and grand totals in the table are denoted as \(N_{i\bullet} = \sum_{j=1}^J N_{ij}\) for \(i=1,\dots, I\); \(N_{\bullet j} = \sum_{i=1}^I N_{ij}\) for \(j=1,\dots, J\); and \(N_{\bullet \bullet} = \sum_{i=1}^I \sum_{j=1}^J N_{ij}\). Furthermore, we denote \(E_{ij}\) as the \textit{null baseline expected count} representing the population-level occurrence of reports for the \((i, j)\) AE-drug pair under the assumption of no association or dependence between the pair \((i, j)\). In practice, \(E_{ij}\) is often estimated/approximated by its natural estimator \(N_{i \bullet} N_{\bullet j} / N_{\bullet \bullet}\), obtained through the marginal row and column proportions \(N_{i \bullet} / N_{\bullet \bullet}\) and \(N_{\bullet j} / N_{\bullet \bullet}\) along with the grand total \(N_{\bullet \bullet}\). Tan et al.\citep{tan2025flexibleempiricalbayesianapproaches} introduced a new estimator, \(\tilde E_{ij} = \frac{N_{iJ}N_{Ij}}{N_{IJ}}\), and established the superiority of \(\tilde E_{ij}\) over the natural estimator \(\hat E_{ij} = N_{i \bullet} N_{\bullet j} / N_{\bullet \bullet}\) under certain settings.

Our base model assumption for the observed count \(N_{ij}\) conditional on \(E_{ij}\) is that
\begin{equation} \label{eqn:model-poisson-likelihood}
N_{ij} \mid E_{ij} \sim \operatorname{Poisson}(E_{ij}\lambda_{ij}),  
\end{equation}
where the parameter \(\lambda_{ij} \geq 0\) is the relative reporting ratio, the signal strength, for the \((i, j)\)-th pair measuring the ratio of the actual expected count arising due to dependence to the null baseline expected count. Therefore, \(\{\lambda_{ij}\}\) are our key parameters of interest. A large \(\lambda_{ij}\) indicates a strong association between a drug and an AE. Formally, an AE-drug combination with a corresponding signal strength \(\lambda > 1\) is defined as a potential signal, and the combination is considered as a non-signal if \(\lambda \leq 1\). In addition, if \(\lambda = 0\), the AE-drug combination is considered impossible to co-occur in the population and is deemed a \textit{structural zero}. This is different from an observed \(N_{ij} = 0\), which may occur with positive probability under the above Poisson law even when \(\lambda_{ij} > 0\). Furthermore, the assumption regarding the baseline AE (\(I\)) and baseline drug level (\(J\)) implies that AE-drug combinations in the last row or last column of the contingency table are necessarily non-signals, with \(\lambda_{iJ} = 1\) for \(i = 1, \dots, I\) and \(\lambda_{Ij} = 1\) for \(j = 1, \dots, J\).

\begin{table}[ht]
\centering
\begin{tabular}{lcc}
\hline
                 & drug-j & Other drugs \\
\hline
AE-i          & $N_{ij}$       & $N_{iJ}$        \\
Other AEs          & $N_{Ij}$       & $N_{IJ}$        \\
\hline
\end{tabular}
\caption{A $2 \times 2$ subtable from the $I \times J$ SRS contingency table.}
\end{table}

Table 2 displays a \(2 \times 2\) sub-contingency table extracted from the full \(I \times J\) SRS contingency table. Using the proposed null value estimator \(\tilde E\), the \(O/E\) ratio for the \((i, j)\)-th AE-drug combination is given by
\[\frac{N_{ij}}{\tilde E_{ij}} = \frac{N_{ij}N_{IJ}}{N_{Ij}N_{iJ}} = \text{OR}_{ij}, \]
where \(\text{OR}_{ij}\) denotes the sample odds ratio derived from the \(2 \times 2\) table in Table 2. This equivalence provides an interpretation of \(\lambda_{ij}\) as the odds ratio associated with the \(2 \times 2\) sub-contingency table formed by AE-\(i\), drug-\(j\), and their reference categories.

\subsection{Non-Parametric Empirical Bayes Models for Pharmacovigilance}\label{non-parametric-empirical-bayes-models-for-pharmacovigilance}

This subsection reviews the KM, Efron and general-gamma approaches within a non-parametric empirical Bayes framework that flexibly estimate \(\{\lambda_{ij}\}\) within the Poisson model \eqref{eqn:model-poisson-likelihood}. Here, the priors for \(\lambda_{ij}\)'s are endowed with a general, non-parametric structure, often in the form of general mixtures of parametric distributions, and a data-driven approach is adopted to estimate the (underlying parameters of the) prior from observed data. In the following, we describe a general structure for non-parametric empirical Bayes methods for the Poisson model \eqref{eqn:model-poisson-likelihood}.

Let \(g\) be a prior density function for signal strength parameters for all AE-drug pairs: \(\lambda_{ij} \sim g\). Then, in the context of the Poisson model \eqref{eqn:model-poisson-likelihood}, the marginal probability mass function of \(N_{ij}\) is given by:
\[
p(N_{ij}) = \int_0^{\infty} g(\lambda_{ij}) \ f_{\text{pois}}(N_{ij} \mid \lambda_{ij}E_{ij}) \ d\lambda_{ij},
\]
where \(f_{\text{pois}}(N \mid \lambda)\) is the probability mass function of a Poisson random variable with mean \(\lambda\) evaluated at \(N\). Under the empirical Bayes framework, the prior distribution is consequently estimated from the data by maximizing the log marginal likelihood:
\[
\hat g = \mathop{\arg \max}\limits_{g} \sum_{i=1}^I \sum_{j=1}^J \log p(N_{ij}).
\]
Then, the estimated posterior density of \(\lambda\) given \(N_{ij}\) is:
\[
\hat{\text{p}}(\lambda \mid N_{ij}) = \frac{\hat g(\lambda) f_{\text{pois}}(N_{ij} \mid \lambda E_{ij})}{\hat{\text{p}}(N_{ij})},
\]
where \(\hat{\text{p}}(N_{ij}) = \int_0^{\infty} \hat g(\lambda_{ij})f_{\text{pois}}(N_{ij}\mid \lambda_{ij}E_{ij}) \ d\lambda_{ij}\).

The above construction fits into the \(g-\)modeling strategy for empirical Bayes estimation discussed by \cite{efron2014two}. It makes no specific assumption about the structure of \(g\). By making \(g\) a flexible non-parametric structure, eg. a form of general mixtures of parametric distributions for model parameters, we may obtain rigorous (empirical) Bayesian inference on \(\{\lambda_{ij}\}\) without the prior \(g\) overshadowing the information provided by the SRS data table while still permitting adequate regularization. This mixture structure of the prior \(g\) allows multiple modes/clusters in the underlying distribution, and thus can flexibly accommodate multiple subgroups of signal, non-signal, or zero-inflation \(\{\lambda_{ij}\}\) values. However, the parametric form for the mixture component densities and the number of mixture components should be carefully determined. Otherwise, posterior inference of \(\{ \lambda_{ij} \}\) may suffer from noisy estimation or underfitting issues due to too small or too large number of mixture components.

\paragraph{KM method}\label{km-method}

The Koenker and Mizera non-parametric empirical Bayes method \citep{koenker2014convex} for Poisson model \eqref{eqn:model-poisson-likelihood} assumes the prior \(g\) for \(\{\lambda_{ij}\}\) to base on a given finite discrete support of size \(K\), \(\lambda \in \{ v_1,...,v_K\}\), \(K<\infty\), with associated prior probability masses: \(\{ g_1,...,g_K\}\) with \(g_k \geq 0\); \(\sum_{k=1}^K g_k = 1\). Under this prior construction, the marginal distribution of an AE-drug pair count is in the form of a mixture of K components Poisson distributions. Then, the model parameters \(\{ g_1,...,g_K\}\) are estimated from maximizing the joint marginal likelihood in the form of convex optimization. The grid values \(\{ v_1,...,v_K\}\) are influential to the model inference, and thus, need to be carefully selected.

\paragraph{Efron's approach}\label{efrons-approach}

Similar to KM, the Efron non-parametric empirical Bayes \citep{efron2016empirical} approach also assumes a finite discrete support of size \(K\): \(\lambda \in \{ v_1, \ldots, v_K \}\), \(K < \infty\). However, instead of assigning arbitrary probabilities, the associated prior probability masses utilize an exponential form:
\[
g = g(\alpha) = \exp\{ Q \alpha - \phi(\alpha) \},
\]
where \(\alpha\) is a \(p\)-dimensional parameter vector , \(Q\) is a known \(K \times p\) structure matrix , and \(\phi > 0\) is an appropriately determined normalizing constant that makes \(g\) a proper mass function. The default choice for \(Q\) in the Efron model is considered to be the basis of a natural spline for \(\{ v_1, \ldots, v_K \}\) with \(p\) degrees of freedom. The prior parameter \(\alpha\) is regularized and estimated through a ridge-penalized log marginal likelihood. Specifically,\\
\[
\hat \alpha = \underset{\alpha}{\arg\max} \log L_{E}(\alpha ; c_0, p) = \underset{\alpha}{\arg\max} \left\{ \sum_{i=1}^I\sum_{j=1}^J \log P_{ij}^T g(\alpha) - c_0\left( \sum_{l=1}^p \alpha_l^2\right)^{1/2} \right\},
\]
where \(P_{ij} = (f_{\text{pois}}(N_{ij} \mid v_k E_{ij}): k= 1, \dots, K)\). Fitting the Efron model requires the specification of hyperparameters \((c_0, p)\). \cite{tan2025flexibleempiricalbayesianapproaches} suggest a AIC-based grid searching approach for choosing \((p, c_0)\). The AIC is defined as follows:
\[ 
    AIC_{\text{E}}(c_0, p) = 2\times \text{trace}(\text{F}) - 2\log L_{E;0}(\hat \alpha),
\]
where \(\text{F} = H_E^{-1}(\hat \alpha ; c_0, p)H_{E}(\hat \alpha ;0,  p)\) is the degrees of freedom matrix\citep{wood2017generalized}.

\paragraph{K-gamma method}\label{k-gamma-method}

As an alternative to the discrete priors used in the KM and Efron approaches, the \(K\)-gamma method uses a continuous prior density \(g\) modeled as a finite mixture of \(K\) gamma distributions, where \(K \geq 3\) is prespecified. The prior takes the form:
\[
g(\lambda\mid R, H, \Omega) = \sum_{k=1}^K \omega_k f_{\text{gamma}} \left(\lambda\mid \alpha = r_k, \beta = \frac{1}{h_k}\right),
\]
where \(\Omega = \{\omega_1, \dots, \omega_K\}\), \(R = \{r_1, \dots, r_K\}\), and \(H = \{h_1, \dots, h_K\}\) denote the component-specific mixture weights, shape, and rate parameters, respectively. Due to the Gamma-Poisson conjugate property, both the marginal and posterior distributions under the \(K\)-gamma model have closed-form expressions: the marginal distribution is a mixture of negative binomials, while the posterior distribution is a mixture of gamma distributions.

Utilizing \(K \geq 3\) gamma components increases the model flexibility, enabling it to better capture complex multi-modal structures in SRS data than simpler variants like the GPS model \citep{dumouchel1999bayesian}, which employs only a \(K=2\) two-component gamma mixture prior. Provided that \(K\) matches with the number of latent subgroups or clusters in the true distribution of the \(\{\lambda_{ij}\}\), this model is capable of appropriately acknowledging the underlying \(\lambda\) heterogeneity. An incorrect \(K\), however, may lead to underfitting or overfitting issues, making the model either too restrictive or too noisy, respectively. In applications, when prior knowledge of K is available, the \(K\)-gamma model can be used. In scenarios where K is unknown, a common case in practice, the general-gamma model, introduced below, provides a more flexible alternative.

\paragraph{General-gamma method}\label{general-gamma-method}

The general-gamma model improves upon the \(K\)-gamma mixture prior model with a fixed \(K\) by adaptively determining the number of mixture components from the data. Specifically, we extend the framework of the \(K\)-gamma mixture prior model to handle situations where no information on the number of components \(K\) is available a priori and thus must be inferred from the data. To achieve this, we adopt the framework of sparse finite mixture models \citep{fruhwirth2019here, malsiner2016model, malsiner2017identifying}, while ensuring stable estimation of the mixture weights. In particular, we begin with an overfitted mixture model---e.g., \(K = 100\) in our implementation---and place a Dirichlet prior on the mixture weights, \(\Omega = \{\omega_1, \dots, \omega_K\} \sim \text{Dirichlet}(\alpha,\alpha, \dots, \alpha)\) with \(0\leq \alpha < 1\). This prior encourages sparse mixture by encouraging many of the mixture weights to shrink toward zero, thereby allowing the model to automatically identify and retain only the most important components. For the hyperparameter \(\alpha = 1\), there is no shrinkage effect of the Dirichlet hyperprior on the resulting posterior distribution. In this case, the fitted model corresponds to either the \(K\)-gamma model or the GPS model, depending on the initialization of \(K\). Conversely, setting \(\alpha = 0\) induces the strongest shrinkage effect, pushing most of the mixture weights toward zero in the posterior. \cite{rousseau2011asymptotic} showed that a Bayesian sparse overfitted mixture (i.e., a mixture fitted with a larger \(K\) than what is actually needed for the data) asymptotically converges to the ``true'\,' population mixture when the Dirichlet prior hyperparameters \(\alpha\) is smaller than \(d/2\), where \(d\) is the dimension of the component-specific parameter \(\{R, H \}\) (\(d = 2\) in general-gamma model). In the \CRANpkg{pvEBayes} package, we implement a bi-level \textbf{ECM algorithm} proposed by \cite{tan2025flexibleempiricalbayesianapproaches} to fit this model. The ECM algorithm is discussed in the following.

Under the Poisson model (1), and conditioning on the mixture weights \(\Omega = \{\omega_1,...,\omega_K\}\), the marginal distribution of \(N_{ij}\) follows a mixture of negative binomial distributions:
\begin{equation}
p(N_{ij} \mid \omega_1, \dots, \omega_K) = \sum_{k = 1}^K \omega_k f_{\text{NB}}\left(N_{ij} \mid r_k, \ \theta_{ijk} = \frac{1}{1 + E_{ij} h_k}\right),
\end{equation}
where \(f_{\text{NB}}(x\mid r, \theta)\) is the probability mass function of a negative binomial distribution with size parameter \(r\) and probability parameter \(\theta\). The bi-level ECM algorithm maximizes the joint marginal likelihood of the general-gamma model by leveraging two distinct data augmentation schemes, one at each level, to facilitate the computations of parameter updates. The first level introduces the latent component indicator \(S = \{S_{11}, S_{12}, \dots, S_{IJ}\}\), \(S_{ij} = (S_{ijk}: k = 1, \dots, K)\), where \(S_{ijk} = 1\) if observation \((i, j)\) is assigned to component \(k\), and 0 otherwise. This augmentation is used to update \(\{ h_k\}\). The second level exploits the Poisson-logarithmic representation of the negative binomial distribution \citep{quenouille1949relation}, introducing latent Poisson random variable \(M\) and associated logarithmic series variables \(Y\), together with \(S\), to facilitate the estimation of \(\{\omega_k, r_k\}\). These two augmentation schemes yield separate expected complete data log-likelihood (``\(Q\)'') functions, which are maximized conditionally in alternating conditional-maximization (CM) steps to update \(\{\omega_k, r_k\}\) and \(\{ h_k\}\), respectively. A common expectation (E) step is employed to update the expected complete data log-likelihood functions in each iteration. The full ECM procedure is outlined in Algorithm 1 below. Detailed derivations and proof of the convergence guarantee are provided in \cite{tan2025flexibleempiricalbayesianapproaches}.

Careful initialization of model parameters is crucial for stable convergence of the ECM algorithm. We begin with an overfitted model by setting a large number of mixture components: \(K = \min\{200, I \times J\}\). To initialize the component-specific parameters, we adopt a ``mean-variance'' strategy. Specifically, we generate a grid of values \(\{v_1, \dots, v_K\}\) and set \(r_k h_k = v_k\) and \(r_k h_k^2 = \epsilon\) for some small \(\epsilon > 0\) (e.g., \(\epsilon = 10^{-6}\)), ensuring each gamma component is centered at \(v_k\) with low variance. The mixture weights are initialized uniformly as \(\omega_k = 1/K\) for all \(k\). The grid \(\{v_1, \dots, v_K\}\) is constructed by drawing randomly from the empirical distribution (histogram) of the observed-to-expected ratios \({N_{ij}/E_{ij}}\), as proposed by \cite{tan2025flexibleempiricalbayesianapproaches}.

The performance of the model can be sensitive to the choice of the Dirichlet hyperparameter \(\alpha\). In particular, smaller values of \(\alpha\) encourage sparser mixtures by reducing the number of non-empty components in the fitted model. To select an appropriate value for \(\alpha\), \cite{tan2025flexibleempiricalbayesianapproaches} suggested using the Akaike Information Criterion (AIC) \citep{akaike1974new}. For a given \(\alpha\), the AIC for the fitted general-gamma model is
\begin{equation}
\text{AIC}(\alpha) = 2\times (3K^*) - 2\log \hat L(\alpha),
\end{equation}
where \(\log \hat L(\alpha) = \log L(\hat \Omega, \hat R, \hat H \mid \alpha)\) is the maximized log-marginal-likelihood (2) obtained in the process of fitting the general gamma model for a given value of \(\alpha\), and \(K^*\) is the number of retained non-empty components in the fitted model. In our package, we also provide functionality for the Bayesian information criterion (BIC) \citep{schwarz1978estimating} to select \(\alpha\). With a given \(\alpha\), the BIC for the fitted general-gamma model is
\begin{equation}
\text{BIC}(\alpha) = (3K^*)\log(I\times J) - 2\log \hat L(\alpha).
\end{equation}

\begin{algorithm}[h] 
\caption{The $(u+1)$-th iteration of the proposed bi-level ECM algorithm for implementation of the general-gamma model.}
\begin{algorithmic}
\STATE \textbf{Require:} Current iteration of $\phi^{(u)} = \{ \Omega^{(u)}, R^{(u)},H^{(u)}\}$. Do for all $k = 1, \dots, K$: \\
\textbf{E step}: Compute $\tau_{ijk}^{(u+1)} = E(\mathbbm{1}_{\{ S_{ijk}=1\}}\mid N,\phi^{(u)})$, $\delta_{ijk}^{(u+1)}=E(M_{ij}\mid S_{ijk}=1,\phi^{(u)})$. \\
\textbf{CM step 1}: Given $h^{(u)}_k$, update $\omega^{(u+1)}_k$ and $r^{(u+1)}_k$ as follows: \\
$$\omega_k^{(u+1)} = \max\left\{0,\frac{\alpha-1 + \sum_{i=1}^I\sum_{j=1}^J\tau_{ijk}^{(u+1)}}{I*J+K(\alpha-1)}\right\}$$
$$r_k^{(u+1)} = \frac{\sum_{i=1}^I\sum_{j=1}^J\tau_{ijk}^{(u+1)}\delta_{ijk}^{(u+1)}}{\sum_{t=1}^T\tau_{ijk}^{(u+1)}\log\theta_{ijk}^{(u)}}$$
\textbf{CM step 2}: Given $r^{(u+1)}_k$ update $h^{(u+1)}_k$ by solving the following equation:
\[
\sum_{i=1}^I\sum_{j=1}^J \tau_{ijk}^{(u+1)}\left[ \frac{N_{ij}}{h_k} - \frac{E_{ij}(N_{ij}+r_k^{(u+1)})}{1+E_{ij} h_k}\right] = 0.
\]
A simple iterative process for solving this equation suggested by Tan et al. (2025) is implemented. 
\end{algorithmic}
\end{algorithm}

\section{\texorpdfstring{A simulation study illustrating the hyperparameter \(\alpha\) selection for the general-gamma method}{A simulation study illustrating the hyperparameter \textbackslash alpha selection for the general-gamma method}}\label{a-simulation-study-illustrating-the-hyperparameter-alpha-selection-for-the-general-gamma-method}

In this section, we explore the signal estimating performance of the general-gamma method with hyperparameter \(\alpha\) selected by AIC and BIC through a simulation. This study extends the work of \cite{tan2025flexibleempiricalbayesianapproaches}, employing the same data generating process, reference table, and evaluation metrics.

\subsection{Data generation}\label{data-generation}

For the data generation, We consider the same reference table, ``statin-42'\,`, and random table generating process (Algorithm 2 provided below) used in \cite{tan2025flexibleempiricalbayesianapproaches}. Algorithm 2 is accessible in our \CRANpkg{pvEBayes} package via generate\_contin\_table(). The reference table has \(I=43\) AEs---including a reference AE ``other AEs''---and \(J=7\) drugs---including a reference drug ``other drugs''. Algorithm 2 generates random tables with a multinomial-based data-generating process, provided with a prespecified matrix of `true' signal strengths \(((\lambda_{ij}^{\text{true}}))\), where we set \(\lambda_{ij}^{\text{true}} > 1\) for all true signal cells, \(\lambda_{ij}^{\text{true}} = 1\) for all true non-signal cells, and \(\lambda_{ij}^{\text{true}} = 0\) for all structural zero cells (if present).

\begin{algorithm}
\caption{Multinomial AE-drug report count data $\{N_{ij}\}$ generation process} \label{alg:multinomial_data_generation}
\begin{algorithmic}
\STATE \textbf{Require:} A signal strength matrix $((\lambda_{ij}: i = 1, \dots, I; j = 1, \dots, J))$, an exemplar dataset $((N_{ij}: i = 1, \dots, I; j = 1, \dots, J))$, and a structural zero probability $\omega$.

\STATE

\STATE 1. Compute the grand total $\tilde N_{\bullet \bullet} = \sum_{i=1}^I\sum_{j=1}^J \tilde N_{ij}$, row totals $\{\tilde N_{i\bullet} = \sum_{j=1}^J \tilde N_{ij}: i = 1, \dots, I\}$, and column totals $\{\tilde N_{\bullet j} = \sum_{i=1}^I \tilde N_{ij}: j = 1, \dots, J\}$. Also compute the corresponding row and column marginal proportions $\{ p_{i \bullet}^* = \tilde N_{i \bullet}/\tilde N_{\bullet \bullet}: i = 1, \dots, I\} $ and $\{p_{\bullet j}^* = \tilde N_{\bullet j}/ \tilde N_{\bullet \bullet}: j = 1, \dots, J \}$.

\STATE 2. Generate structural zero position indicators $z_{ij} \sim \text{Bernoulli}(\omega)$ such that $z_{ij} = 1$ implies cell $(i, j)$ is a structural zero.

\STATE 3. Compute the cell probabilities $P = (p_{11}, p_{12}, \dots, p_{IJ})$ such that:
\[
p_{ij} = \frac{(1-z_{ij})\lambda_{ij}p_{i\bullet}^*p_{\bullet j}^*}{\sum_{i=1}^I\sum_{j=1}^J (1-z_{ij})\lambda_{ij}p_{i\bullet}^*p_{\bullet j}^*}.
\]

\STATE 4. Generate a random table $((N_{ij}))$ with $(N_{11}, N_{12}, \dots, N_{IJ}) \sim \text{Multinomial}(\tilde N_{\bullet \bullet}, P)$.
\end{algorithmic}
\end{algorithm}

For the simulation setting configurations, we consider a homogeneous signal strength setting, where all 3 signal cells \(\{(i, j)\}\) are assigned a common value as their true signal strength \({\lambda_{ij}^{\text{true}}}\) that is varied over \(\{1.2, 1.4, 1.6, 2.0, 2.5, 3.0, 4.0\}\). For each \(\left((\lambda_{ij}^{\text{true}})\right)\) matrix generated as described above, we considered two structural zero-inflation settings: (1) no zero-inflation, where the \({\lambda_{ij}^{\text{true}}}\) values remained unchanged, (2) moderate zero-inflation, where 25\% of the \({\lambda_{ij}^{\text{true}}}\) values were randomly set to zero, representing true structural zeros, and (3) high zero-inflation, where 50\% of the \({\lambda_{ij}^{\text{true}}}\) values were randomly set to zero. In (2) and (3), these structural zeros were assigned randomly among the non-signal cells (i.e., those with \(\lambda_{ij}^{\text{true}} = 1\)), excluding the reference row \(I\) and reference column \(J\).

\begin{table}[ht]
\centering
\begin{tabular}{lll}
\hline
Factors        & Level & Number of levels \\ \hline
Number of signal cells &  3; $\{(1,1), (7,1), (9,1)\}$           & 1 \\
Signal strength & 1.2, 1.4, 1.6, 2.0, 2.5, 3.0, 4.0 & 7 \\
Level of zero-inflation  & None, Moderate, High          & 3 \\\hline
\end{tabular}
\caption{Random table configurations in the simulation (homogeneous signal strengths). The number of signal cells is followed by the positions of each signal cell}
\end{table}

\subsection{Assessment}\label{assessment}

The signal estimating metrics considered in this simulation are Max-Scaled-RMSE and Average-Scaled-RMSE suggested by \cite{tan2025flexibleempiricalbayesianapproaches}, which measure the deviation between the posterior distribution and the true signal strength.

Let \(f_{ij}\) denote the posterior density of the signal strength parameter \(\lambda_{ij}\) given an SRS dataset. To derive evaluation metrics for \(f_{ij}\) to be used as a Bayesian estimator of the signal strength parameter \(\lambda_{ij}\), we consider the general scaled \(p\)-th Wasserstein distance between \(f_{ij}\) and the point mass at \(\lambda_{ij}^{\text{true}}\), which is defined as:
\[
\text{Scaled-Wasserstein}_p(f_{ij}, \lambda_{ij}^{\text{true}}) = \frac{1}{\lambda_{ij}^{\text{true}}}\left[\int_{0}^{1} \left|F_{ij}^{-1}(q)-F_{\lambda_{ij}^{\text{true}}}^{-1}(q)\right|^p dq\right]^{1/p},
\]
where \(F_{ij}\) and \(F_{\lambda_{ij}^{\text{true}}}\) are the cumulative distribution functions associated with \(f_{ij}\) and the degenerate distribution characterized by a point-mass at \(\lambda_{ij}^{\text{true}}\), respectively, and \(F_{ij}^{-1}\) and \(F_{\lambda_{ij}^{\text{true}}}^{-1}\) are the corresponding quantile functions. To compute these Wasserstein distances,
\cite{tan2025flexibleempiricalbayesianapproaches} exploit the degeneracy of \(F_{\lambda_{ij}^{\text{true}}}\), allowing to use an equivalent formulation expressed as an expectation with respect to the posterior density \(f_{ij}\):
\[
\text{Scaled-Wasserstein}_p(f_{ij}, \lambda_{ij}^{\text{true}}) =
\frac{1}{\lambda_{ij}^{\text{true}}}\left[\int_{0}^{\infty} \left|\lambda-\lambda_{ij}^{\text{true}}\right|^p f_{ij}(\lambda)d\lambda\right]^{1/p}.
\]
In this simulation, we focus on the case \(p=2\) and evaluate these distances by posterior draws. Let \(\lambda_{ijm}^{(s)}\) denote the \(s\)-th posterior draws for \(i\)-th AE and \(j\)-th drug in \(m\)-th replicated table, where \(s = 1,\dots, S\); \(i=1,\dots, I\); \(j=1,\dots, J\); \(m = 1,\dots, M\). Then, the scaled Wasserstein-2 distance between \(f_{ij}\) and \(\lambda_{ij}^{\text{true}}\) is evaluated as:
\[
 \text{Scaled-Wasserstein}_2(f_{ij}, \lambda_{ij}^{\text{true}}) \approx \left[ \frac{1}{S} \sum_{s=1}^S \left( \frac{\lambda_{ijm}^{(s)} - \lambda_{ij}^{\text{true}}}{\lambda_{ij}^{\text{true}}}\right)^2\right]^{1/2}
\]
Closed-form expressions for the resulting distances with \(p = 1\) and \(p = 2\) can be derived when the posterior distribution \(f_{ij}\) has a gamma-mixture form \citep{tan2025flexibleempiricalbayesianapproaches}. In our simulations, we set \(S = 10,000\) to ensure sufficiently accurate Monte Carlo evaluation of these distances. The Wasserstein distance for \(p = 2\) is hereafter referred to as the scaled posterior root mean squared error (scaled posterior RMSE). Based on these distances, we define the following two evaluation metrics that aggregate performance across all signal cells in the table:
\begin{equation*}
    \begin{split}
        & \text{Average-Scaled-Wasserstein}_p(\{f_{ij}\}, \{\lambda_{ij}^{\text{true}}\}) = \frac{1}{\# C_{\text{sig}}} \sum_{(i,j)\in C_{\text{sig}}} \text{Scaled-Wasserstein}_p(f_{ij}, \lambda_{ij}^{\text{true}}), \\
        & \text{Max-Scaled-Wasserstein}_p(\{f_{ij}\}, \{\lambda_{ij}^{\text{true}}\}) = \max_{(i,j)\in C_{\text{sig}}} \left[ \text{Scaled-Wasserstein}_p(f_{ij}, \lambda_{ij}^{\text{true}}) \right],
    \end{split}
\end{equation*}
where \(C_{\text{sig}} = \{(i, j): \lambda_{ij}^{\text{true}} > 1 \}\) is the set of all \textit{true} signal cells, and \(\#C_{\text{sig}}\) denotes its cardinality. These metrics are used to summarize the signal strength estimation performance of a method on a given simulated dataset. To assess overall performance across all replicates, we compute replication-based (i.e., frequentist) averages of these distances. Specifically, given the estimated posterior densities \(f_{ij}^{(m)}\) from the \(m\)-th replicate for \(m = 1, \dots, M = 1000\), we compute the following summary metrics:
\begin{equation} 
    \begin{split}
        &\text{Average-Scaled-RMSE} = \frac{1}{M}\sum_{m=1}^M\text{Average-Scaled-Wasserstein}_2(\{f_{ij}^{(m)}\}, \{\lambda_{ij}^{\text{true}}\}), \\
        &\text{Max-Scaled-RMSE} = \frac{1}{M}\sum_{m=1}^M\text{Max-Scaled-Wasserstein}_2(\{f_{ij}^{(m)}\}, \{\lambda_{ij}^{\text{true}}\}).
    \end{split}
\end{equation}
We use these metrics to assess the estimating performance of the general-gamma model with the hyperparameter \(\alpha\) selected by different methods.

\subsection{Hyperparameter selection methods}\label{hyperparameter-selection-methods}

In this simulation, We consider 5 different hyperparameter selection methods, which are listed below:

\begin{itemize}
\item fix-0: $\alpha$ is fixed at 0.
\item fix-0.5: $\alpha$ is fixed at 0.5.
\item fix-0.9: $\alpha$ is fixed at 0.9.
\item AIC: $\alpha$ is selected by evaluating the AIC for a range of candidate values $(\alpha \in \{0, 0.3, 0.5, 0.7, 0.9\})$.
\item BIC: $\alpha$ is selected by evaluating the BIC for a range of candidate values $(\alpha \in \{0, 0.3, 0.5, 0.7, 0.9\})$.
\end{itemize}

We consider the fixed-\(\alpha\) settings (fix-0, fix-0.5, fix-0.9) as baseline approaches that do not involve hyperparameter tuning. In contrast, the AIC- and BIC-based methods aim to adaptively select \(\alpha\) based on model fit to the observed data. In the following subsections, we evaluate and compare the signal strength estimation performance of general-gamma with the hyperparameter \(\alpha\) selected by AIC and BIC in contrast to the fixed baselines.

\subsection{\texorpdfstring{Conducting the simulation with \CRANpkg{pvEBayes}}{Conducting the simulation with }}\label{conducting-the-simulation-with}

We first define a main simulation function, simulate\_fit\_nsim(), which generates a random SRS contigency table, fits the general-gamma model over the above range of \(\alpha\) values and computes the metrics, Average-Scaled-RMSE and Max-Scaled-RMSE, given the true signal \(\lambda_{ij}^{\text{true}}\), the level of zero inflation, and the number of Monte Carlo replications. We set n\_sim = 200 for this simulation. A larger number of Monte Carlo replications (e.g., n\_sim = 1000) may be used to obtain more accurate results. We use sequential computing for simplicity; however, the computation-heavy steps can be readily parallelized using modern parallel computing frameworks in R, e.g., future-based computing via R packages future.apply and furrr (see the commented codes). The parallelization can be particularly cost-effective when a larger \texttt{n\_sim} is used. The main function is defined below:

\begin{verbatim}
library(pvEBayes)
library(dplyr)
library(tidyr)
library(ggplot2)
set.seed(1)
n_sim <- 200
ref_table <- statin42 # reference table: statin42
simulate_fit_nsim <- function(lambda, zi = 0.25, n_sim, ref_table) {
  
  # generate signal strength matrix
  
  signal_mat <- matrix(1, nrow(ref_table), ncol(ref_table))
  signal_mat[c(1, 7, 9), 1] <- lambda
  E_mat <- estimate_null_expected_count(ref_table)
  zi_indic_mat <- (E_mat <= quantile(E_mat, zi))
  
  # random table generation
  
  list_statin_random <- generate_contin_table(
    n_table = n_sim,
    ref_table = ref_table,
    signal_mat = signal_mat,
    zi_indic_mat = zi_indic_mat
  )
  
  # for each generated table:
  # fit the general-gamma model with different hyperparameter alpha selection
  # compute evaluation metrics for each fitted model
  ## can be parallelized via future.apply::future_lapply (or furrr::future_map)

  metrics_calculation <- lapply(
    list_statin_random,
    function(dat_random) {
      suppressMessages({
        fit_tune <- pvEBayes_tune(dat_random,
                                  model = "general-gamma",
                                  alpha_vec = c(0, 0.3, 0.5, 0.7, 0.9),
                                  return_all_fit = TRUE
        )
      })
      all_fit_objects <- extract_all_fitted_models(fit_tune)
      fit_list <- with(
        all_fit_objects,
        list(
          AIC = best_model_AIC,
          BIC = best_model_BIC,
          fix_0 = all_fit[[1]],
          fix_0.5 = all_fit[[3]],
          fix_0.9 = all_fit[[5]]
        )
      ) %>%
        lapply(
          function(fit) {
            # extract posterior draws for signal cells
            post_draws <- posterior_draws(fit,
                                          n_posterior_draws = 10000
            )$posterior_draws[, c(1, 7, 9), 1]
            
            RMSE_signal <- ((post_draws / lambda - 1)^2) %>%
              apply(c(2), mean) %>%
              sqrt()
            c(
              max_scaled_rmse = max(RMSE_signal),
              ave_scaled_rmse = mean(RMSE_signal)
            )
          }
        ) %>%
        do.call(rbind, .) %>%
        tibble::as_tibble(rownames = "model")
    }
  ) %>%
    do.call(rbind, .) %>%
    group_by(model) %>%
    summarise(
      Max_Scaled_RMSE = mean(max_scaled_rmse),
      Ave_Scaled_RMSE = mean(ave_scaled_rmse)
    ) %>%
    pivot_longer(
      cols = c(Max_Scaled_RMSE, Ave_Scaled_RMSE),
      names_to = "metric_name",
      values_to = "RMSE"
    )
  return(metrics_calculation)
}
\end{verbatim}

Next, we create a grid of simulation setting configurations listing different zero-inflation and signal strength values (created using tidyr::expand\_grid()) and iterate the above simulation and computation function simulate\_fit\_nsim() over the grid using mapply(). Finally, the results are visualized using the \CRANpkg{ggplot2} package. The code is provided below:

\begin{verbatim}
simulation_result <- 
  # create a grid of simulation settings over zero-inflation levels
  # and signal strength values
  tibble(
    zi = c(0, 0.25, 0.50), 
    zi_label = c("ZI:none", "ZI:moderate", "ZI:high") %>%
      factor(levels = unique(.))
  ) %>%
  expand_grid(lambda = c(1.2, 1.4, 1.6, 2, 2.5, 3, 4)) %>% 
  
  # iterate over all simulation configurations and 
  # compute the evaluation metrics
  
  mutate(
    result = mapply(
      simulate_fit_nsim,
      lambda = lambda,
      zi = zi,
      MoreArgs = list(n_sim = n_sim, ref_table = ref_table),
      SIMPLIFY = FALSE
    )
  ) %>% 
  unnest(result)

# simulation result visualization

simulation_plot <- simulation_result %>%
  ggplot(
    aes(
      x = lambda,
      y = RMSE,
      color = model,
      shape = model,
      group = model
    )
  ) +
  geom_point(position = position_dodge(0.03), size = 1.5) +
  geom_line(position = position_dodge(0.03), alpha = 0.4, linewidth = .25) +
  facet_grid(metric_name ~ zi_label) +
  scale_color_manual(values = c(
    "#E69F00", "#56B4E9", "#009E73", "#F0E442", "#D55E00"
  )) +
  scale_shape_manual(values = c(4, 15, 16, 16, 16)) +
  scale_x_continuous(trans = "log10", breaks = c(1.2, 1.4, 1.6, 2, 2.5, 3, 4)) +
  labs(x = "signal strength", y = "Metric value", color = "model") +
  theme_bw() +
  guides(shape = "none") +
  theme(
    legend.position = "top"
  )

simulation_plot + theme(
  axis.text = element_text(size = 13, face = "bold"),
  legend.title = element_text(size = 14, face = "bold"),
  strip.text = element_text(size = 14, face = "bold"),
  axis.title.x = element_text(size = 14),
  axis.title.y = element_text(size = 14),
  legend.text = element_text(size = 14)
)
\end{verbatim}

\begin{center}\includegraphics[width=1\linewidth]{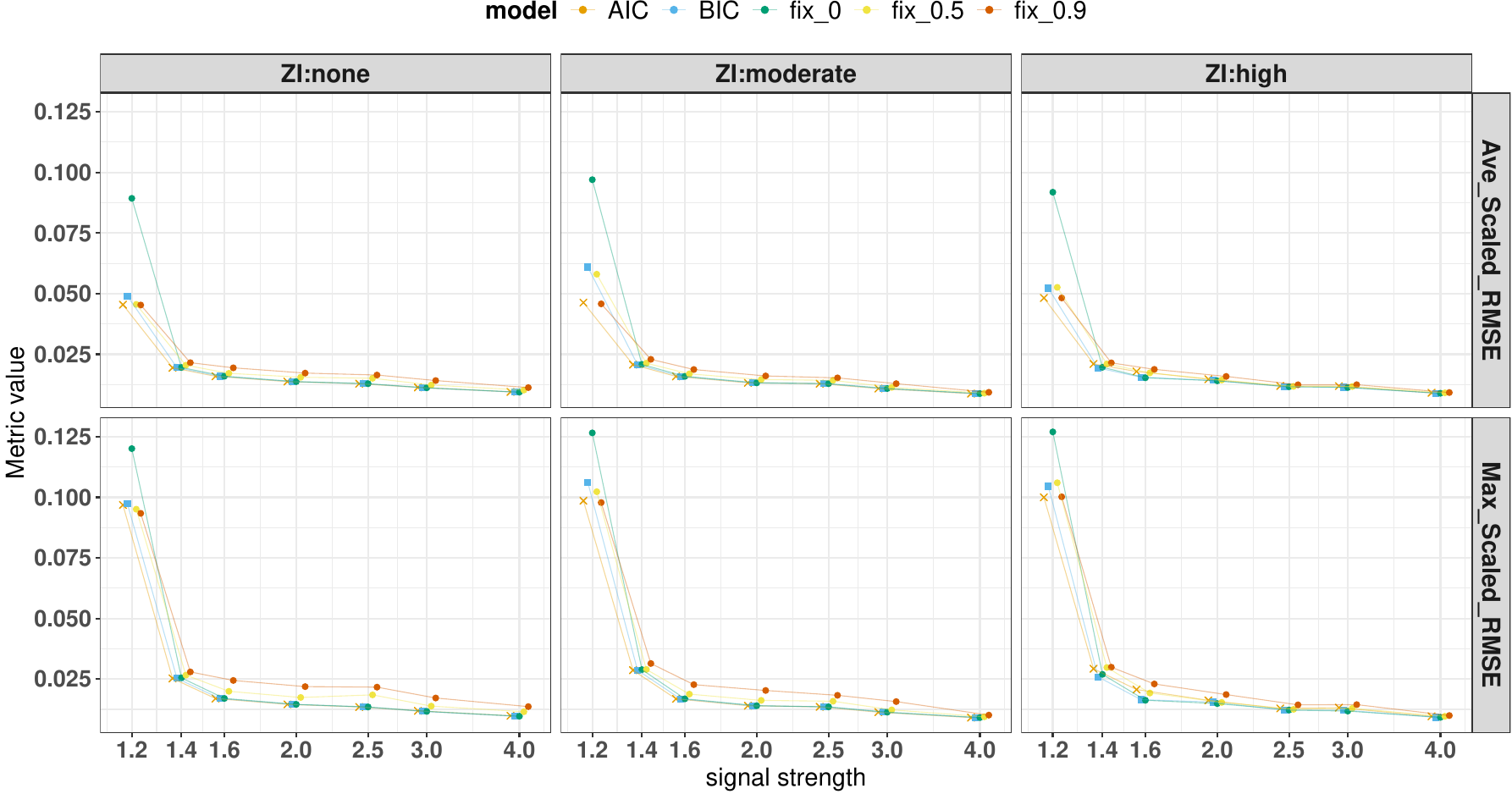} \end{center}

The above figure provides the signal estimating performance of the general-gamma model with hyperparameter \(\alpha\) selected by various hyperparameter selection methods (color-coded). The row panels show the signal estimation metrics, Max-Scaled-RMSE and Average-Scaled-RMSE, respectively, along the vertical axes against different \(\lambda_{ij}^{\text{true}}\) values for the signal cells, across different zero-inflation levels. Dot shapes for AIC and BIC are highlighted via crosses and rectangles.

The estimation performance of the fixed hyperparameter baselines (fix\_0, fix\_0.5, fix\_0.9) illustrates how the choice of \(\alpha\) affects the posterior signal strength estimates across varying values of \(\lambda_{ij}^{\text{true}}\) for signal cells. At the lowest signal level (\(\lambda_{ij}^{\text{true}} = 1.2\)), the estimating performance from the worst to the best are \texttt{fixed\_0}, \texttt{fixed\_0.5}, and \texttt{fixed\_0.9}. As signal strength increases, their order switches. In the general-gamma model, as \(\alpha\) approaches zero, the estimated prior becomes increasingly sparse, resulting in fewer non-empty mixture components in the posterior. When signal strength is weak (i.e., \(\lambda_{ij}^{\text{true}}\) close to 1), small values of \(\alpha\) may lead to over-shrinkage, forcing signal-related components toward empty. As \(\lambda_{ij}^{\text{true}}\) increases, this risk diminishes. However, a large \(\alpha\) value makes Bayesian inference noisy. In addition, the results of fixed hyperparameter baselines indicate that our general-gamma model is robust in the choice of hyperparameter \(\alpha\). Both AIC- and BIC-based selection methods perform well in this simulation, suggesting that they are effective and reasonable strategies for tuning \(\alpha\) in the general-gamma model.

\section{\texorpdfstring{Real data analysis: Analyzing FDA FAERS data with \CRANpkg{pvEBayes}}{Real data analysis: Analyzing FDA FAERS data with }}\label{real-data-analysis-analyzing-fda-faers-data-with}

In this section, we present data analyses using \CRANpkg{pvEBayes} package based on two real-world datasets, ``statin2025'' and ``gbca2025'', both included as contingency tables in the package. Each dataset compiles adverse event (AE) reports from the FDA FAERS database, covering the period from 2020Q1 to 2024Q4, for statin drugs and gadolinium-based contrast agent (GBCA) drugs, respectively. A brief description of each dataset is provided below.

\begin{itemize}
\item Statins are widely prescribed medications that help lower blood levels of low-density lipoprotein (LDL) cholesterol, which is associated with an increased risk of cardiovascular disease. The statin2025 dataset has each entry representing the count of AE reports associated with one of seven drug columns---six statin drugs and a reference group labeled as "other drugs"---across 5,119 AE rows. The statin drugs included in this dataset are: Atorvastatin, Fluvastatin, Lovastatin, Pravastatin, Rosuvastatin, and Simvastatin. 

\item Gadolinium-based contrast agents (GBCAs) are injected drugs to improve image quality in diagnostic procedures such as magnetic resonance imaging (MRI) and magnetic resonance angiography (MRA). The gbca2025 dataset represents AE-drug reports associated with eight drug columns, including seven GBCAs and a reference drug ``other drugs'', across 1,328 AE rows. The GBCAs under consideration in the dataset are: Gadobenate, Gadobutrol, Gadodiamide, Gadopentetate, Gadoterate, Gadoteridol, and Gadoxetate.
\end{itemize}

These datasets extend the original statin and gbca datasets from the \CRANpkg{pvLRT} package. While the datasets in \CRANpkg{pvLRT} cover AE reports from the FAERS database for the period 2014Q3 to 2020Q4, the statin2025 and gbca2025 datasets include more recent AE reports, spanning 2021Q1 to 2024Q4.

\paragraph{Aim of the data analysis}\label{aim-of-the-data-analysis}

Statin drugs are a group of commonly prescribed medications worldwide and there are many studies on their adverse effects. Myalgia and rhabdomyolysis are the most known adverse effects of statin drugs \citep{ramkumar2016statin}. We construct a reduced version of the statin2025 dataset by collapsing AE rows based on a list of adverse events identified as significant by the FDA. This results in a \(45 \times 7\) contingency table, where the rows consist of 44 AEs of interest and one reference row, labeled ``other AEs'', which aggregates all remaining AEs from the original statin2025 dataset. We refer to this collapsed dataset as statin2025\_44. Therefore, statin2025\_44 is likely to contain many signals. We would like to see if the empirical Bayesian approaches implemented in our package detect these known AEs for statin drugs and if yes, what are the estimated signal strengths \((\lambda)\) for these AE-drug combinations.

Gadolinium-based contrast agents (GBCAs) have been widely prescribed since their introduction in 1987. However, over the past decade, growing evidence has raised concerns about the safety of GBCAs, particularly regarding gadolinium deposition in brain tissue \citep{gulani2017gadolinium}. The potential toxicity of gadolinium has been studied by \cite{ramalho2016gadolinium}. In 2017, the U.S. Food and Drug Administration (FDA) required new class warnings for all GBCAs due to the retention of gadolinium in the body---including the brain---for months or even years after administration. Despite these warnings, the FDA concluded that no harmful effects related to gadolinium retention have been identified in patients with normal kidney function and that the benefits of approved GBCAs outweigh the potential risks \citep{fda2017gbca}. To investigate potential brain- and nervous system-related adverse events (AEs) associated with GBCAs, we also constructed a reduced version of the gbca2025 dataset by collapsing AE rows based on a predefined list of neurologically relevant events. The resulting dataset, gbca2025\_69, is a \(70 \times 9\) contingency table, where the rows comprise 69 AEs of interest and one reference row labeled ``other AEs'', which aggregates all remaining AEs from the original gbca2025 dataset.

\subsection{Analysis of the statin2025\_44 data set}\label{analysis-of-the-statin2025_44-data-set}

To begin the analysis, we first load the data into an R session and view the first 6 rows and 3 columns with the following codes:

\begin{verbatim}
data("statin2025_44")
head(statin2025_44)[, 1:4]
\end{verbatim}

\begin{verbatim}
#>                                        Atorvastatin Fluvastatin Lovastatin
#> ACUTE KIDNEY INJURY                            1132          23         23
#> ANURIA                                           46           0          0
#> BLOOD CALCIUM DECREASED                          51           2          0
#> BLOOD CREATINE PHOSPHOKINASE ABNORMAL            19           0          0
#> BLOOD CREATINE PHOSPHOKINASE INCREASED          624          21          4
#> BLOOD CREATININE ABNORMAL                        11           0          0
#>                                        Pravastatin
#> ACUTE KIDNEY INJURY                            153
#> ANURIA                                           1
#> BLOOD CALCIUM DECREASED                          3
#> BLOOD CREATINE PHOSPHOKINASE ABNORMAL            0
#> BLOOD CREATINE PHOSPHOKINASE INCREASED          41
#> BLOOD CREATININE ABNORMAL                        0
\end{verbatim}

Our interest lies in finding the most important adverse events and estimating the corresponding signal strength of these 6 statin drugs. These are achieved by fitting empirical Bayes models with pvEBayes() to the SRS data.

\paragraph{Analysis based on general-gamma model}\label{analysis-based-on-general-gamma-model}

We begin by fitting the general-gamma model to this dataset. For illustration, we fit the model with a fixed hyperparameter value of \(\alpha = 0.5\) using the pvEBayes() function. The first argument specifies the SRS data to be analyzed, and the second argument indicates the model type. The hyperparameter \(\alpha = 0.5\) is supplied as the third argument. The fourth argument, n\_posterior\_draws = 1000, sets the number of (empirical Bayes) posterior draws \({\lambda_{ij}^{(s)}}\) to be generated for each AE-drug combination. Although both the gamma mixture models (GPS, K-gamma, general-gamma) and discrete models (KM, Efron) yield closed-form posterior distributions for each AE-drug pair, posterior draws offer a unified structure for computation and visualization, which is helpful for practical implementation.

\begin{verbatim}
gg_given_alpha <- pvEBayes(statin2025_44,
  model = "general-gamma",
  alpha = 0.5,
  n_posterior_draws = 1000
)
\end{verbatim}

Running the above code returns an object of class pvEBayes, an S3 object (list) that contains the estimated parameters of the gamma mixture prior, information of the model fitting process, posterior draws stored as a 3-d array, and the estimated null expected counts \({\tilde E_{ij}}\). The print method for pvEBayes objects provides a textual summary of the fitted model. More detailed information can be extracted using the summary method:

\begin{verbatim}
gg_given_alpha

gg_given_alpha_detected_signal <- summary(gg_given_alpha,
  return = "detected signal"
)
sum(gg_given_alpha_detected_signal)
\end{verbatim}

\begin{verbatim}
#> [1] 106
\end{verbatim}

The return argument specifying which component the summary function should return. Valid options include: ``prior parameters'', ``likelihood'', ``detected signal'' and ``posterior draws''. If it is set to NULL (default), all components will be returned in a list. We do not show all summary results for brevity here. The object gg\_given\_alpha\_detected\_signal is a matrix of the same size as the SRS data under analysis with each entry indicating whether the corresponding AE-drug combination is a detect signal, which is defined as whether \(\text{Pr}(\lambda_{ij} \geq \text{cutoff} | \text{data}) > 0.95\). The cutoff is greater than 1 and defaults to 1.001. The above computation shows that 106 AE-drug combinations are detected as a signal. In this package we suggest tuning the general-gamma model by AIC or BIC that can be accessed through AIC() or BIC() functions as shown below:

\begin{verbatim}
AIC(gg_given_alpha)
\end{verbatim}

\begin{verbatim}
#> [1] 3802.753
\end{verbatim}

\begin{verbatim}
BIC(gg_given_alpha)
\end{verbatim}

\begin{verbatim}
#> [1] 4016.649
\end{verbatim}

In practice, one can specify a list of candidate \(\alpha\) values, fit the general-gamma model for each, compute the corresponding AIC or BIC, and select the model with the lowest AIC or BIC. Instead of manually doing so, one can use the pvEBayes\_tune() function, which implement these steps. The relevant code is given below:

\begin{verbatim}
gg_tune_statin44 <- pvEBayes_tune(statin2025_44,
  model = "general-gamma",
  alpha_vec = c(0, 0.1, 0.3, 0.5, 0.7, 0.9),
  use_AIC = TRUE,
  n_posterior_draws = 10000
)
\end{verbatim}

\begin{verbatim}
#> The alpha value selected under AIC is 0.3,
#> The alpha value selected under BIC is 0.1.
\end{verbatim}

\begin{verbatim}
#>   alpha      AIC      BIC num_mixture
#> 1   0.0 4551.612 4697.962          13
#> 2   0.1 3799.011 3990.392          17
#> 3   0.3 3798.874 4001.513          18
#> 4   0.5 3802.753 4016.649          19
#> 5   0.7 3822.367 4081.295          23
#> 6   0.9 3906.526 4323.061          37
\end{verbatim}

The input data, model specification, and the number of posterior draws arguments are the same as pvEBayes(). The third argument specifies a vector of candidate \(\alpha\) values and the fourth argument, use\_AIC = FALSE, indicates the model in the pvEBayes object returned by this function is chosen under BIC. Running pvEBayes\_tune() also prints out a table of the tuning process showing AIC and BIC for models with each candidate \(\alpha\) value. The selected hyperparameter values for \(\alpha\) are 0.3 and 0.1, as determined by AIC and BIC, respectively. For the subsequent data analysis, we follow the BIC recommendation where \(\alpha = 0.1\). As shown in the simulation results (Section 5), the performance of the general-gamma model is robust to the choice of \(\alpha\), with values 0.3 and 0.1 leading to similar posterior distributions. Again, we print and summarize the chosen model.

\begin{verbatim}
gg_tune_statin44

gg_tune_statin44_detected_signal <- summary(gg_tune_statin44,
  return = "detected signal"
)
sum(gg_tune_statin44_detected_signal)
\end{verbatim}

\begin{verbatim}
#> [1] 104
\end{verbatim}

The above computation shows that 104 AE-drug combinations are detected as a signal which is similar to the general-gamma model with a given hyperparameter \(\alpha = 0.5\). In addition, \CRANpkg{pvEBayes} has implemented visual summary methods for both signal detection and estimation, which are heatmap and eyeplot. These plot functions can be accessed through the plot() with argument type = ``heatmap'' or type = ``eyeplot''. In the following, we create a heatmap for the top 10 significant AEs across all 6 statin drugs. The order of AE rows is determined by the largest \(\text{Scaled-Wasserstein}_2(\hat f_{ij}, 1)\) of each AE row, where \(\hat f_{ij}\) is the estimated posterior distribution of \((i, j)\)-th AE-drug pair. The order of drug columns is determined by the number of detected signals for each drug. Plotting for specific AEs and drugs can be achieved by using AE\_names or drug\_names instead. The heatmap is stored in heatmap\_gg\_statin44.

\begin{verbatim}
heatmap_gg_tune_statin44 <- plot(gg_tune_statin44,
  type = "heatmap",
  num_top_AEs = 10,
  cutoff_signal = 1.001
)
\end{verbatim}

The arguments used in the plot() above are described as follows. The first argument is a pvEBayes object. The second argument is the type of the plot to be made. The third argument represents the number of most significant AEs under consideration. The last argument cutoff\_signal is the threshold for being detected as a signal as defined above.

\begin{verbatim}
heatmap_gg_tune_statin44 +
  theme(
    axis.text = element_text(size = 13, face = "bold"),
    legend.title = element_text(size = 14, face = "bold"),
    strip.text = element_text(size = 14, face = "bold"),
    legend.position = "top"
  )
\end{verbatim}

\begin{center}\includegraphics[width=1\linewidth]{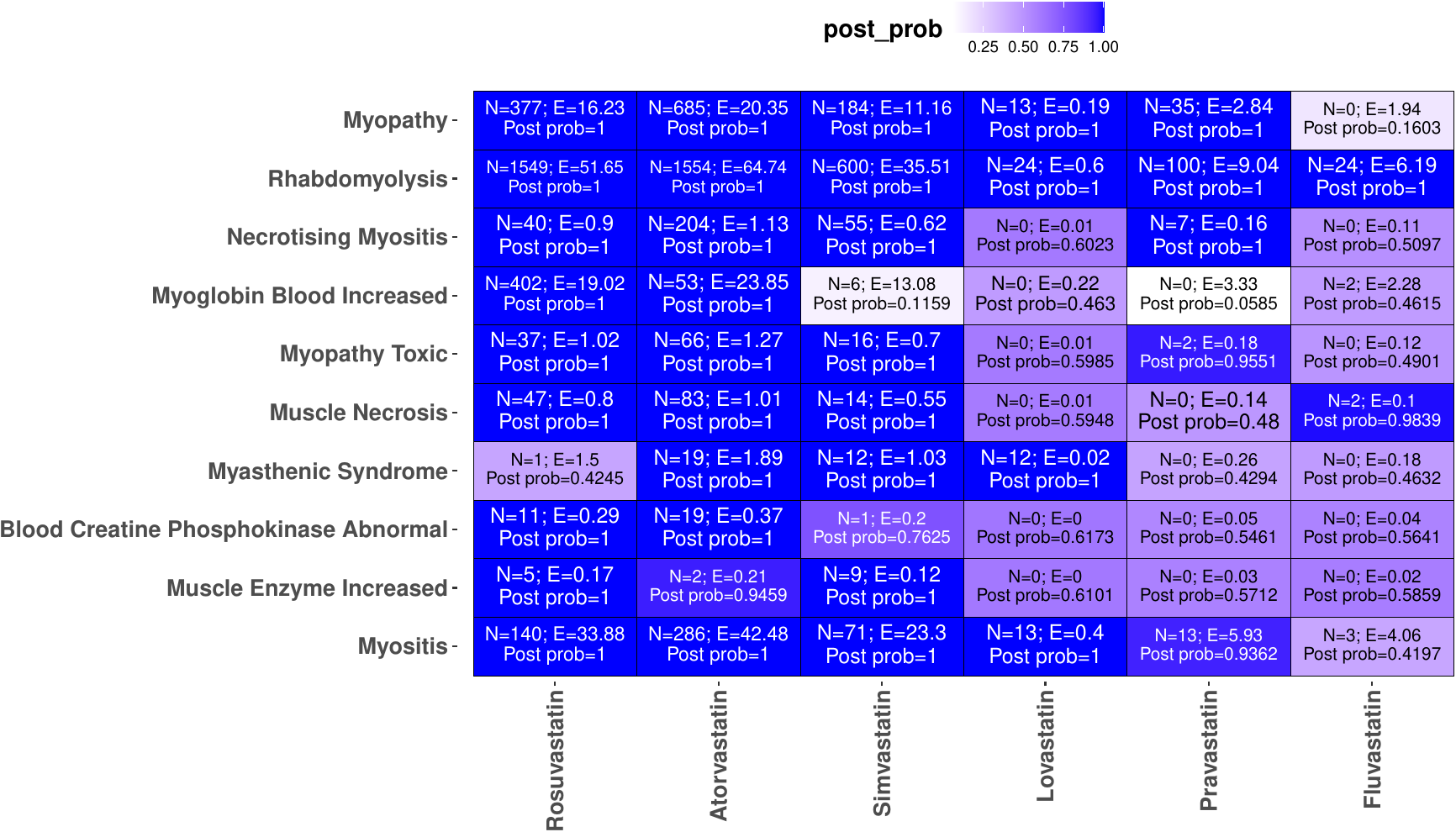} \end{center}

The above heatmap visualizes the signal detection result for the fitted general-gamma model. For each AE-drug combination, the number of reports (N), estimated null value (E), and estimated posterior probability of being a signal are given. Deeper blue color indicates stronger evidence for a signal. To visualize the posterior distribution of signal strength \((\lambda)\) for each AE-drug pair, type = ``eyeplot'' can be used. The following R codes generate the eyeplot. The first three arguments are the same as the heatmap. The argument N\_threshold = 1 specifies that any AE-drug combination with an observed count less than 1 will be excluded from the analysis because the combinations with near-zero observations are unlikely to provide meaningful evidence for or against a signal and often reflect random noise in SRS data. The argument log\_scale = FALSE indicates that the eyeplot will display the posterior distribution of \(\{\lambda_{ij}\}\) for selected AEs and drugs on the original scale, rather than on the logarithmic scale. The rest of the arguments text\_shift,text\_size, and x\_lim\_scalar control the relative position and size of text labels as well as the x-axis range, respectively.

\begin{verbatim}
eyeplot_gg_tune_statin44 <- plot(gg_tune_statin44,
  type = "eyeplot",
  num_top_AEs = 8,
  N_threshold = 1,
  log_scale = FALSE,
  text_shift = 2.1,
  text_size = 4,
  x_lim_scalar = 1.2
) +
  guides(color = guide_legend(nrow = 1)) +
  theme(
    axis.text = element_text(size = 13, face = "bold"),
    legend.title = element_text(size = 14, face = "bold"),
    strip.text = element_text(size = 14, face = "bold"),
    axis.title.x = element_text(size = 14),
    axis.title.y = element_text(size = 14),
    legend.text = element_text(size = 14),
    legend.position = "top"
  )

eyeplot_gg_tune_statin44
\end{verbatim}

\begin{center}\includegraphics[width=1\linewidth]{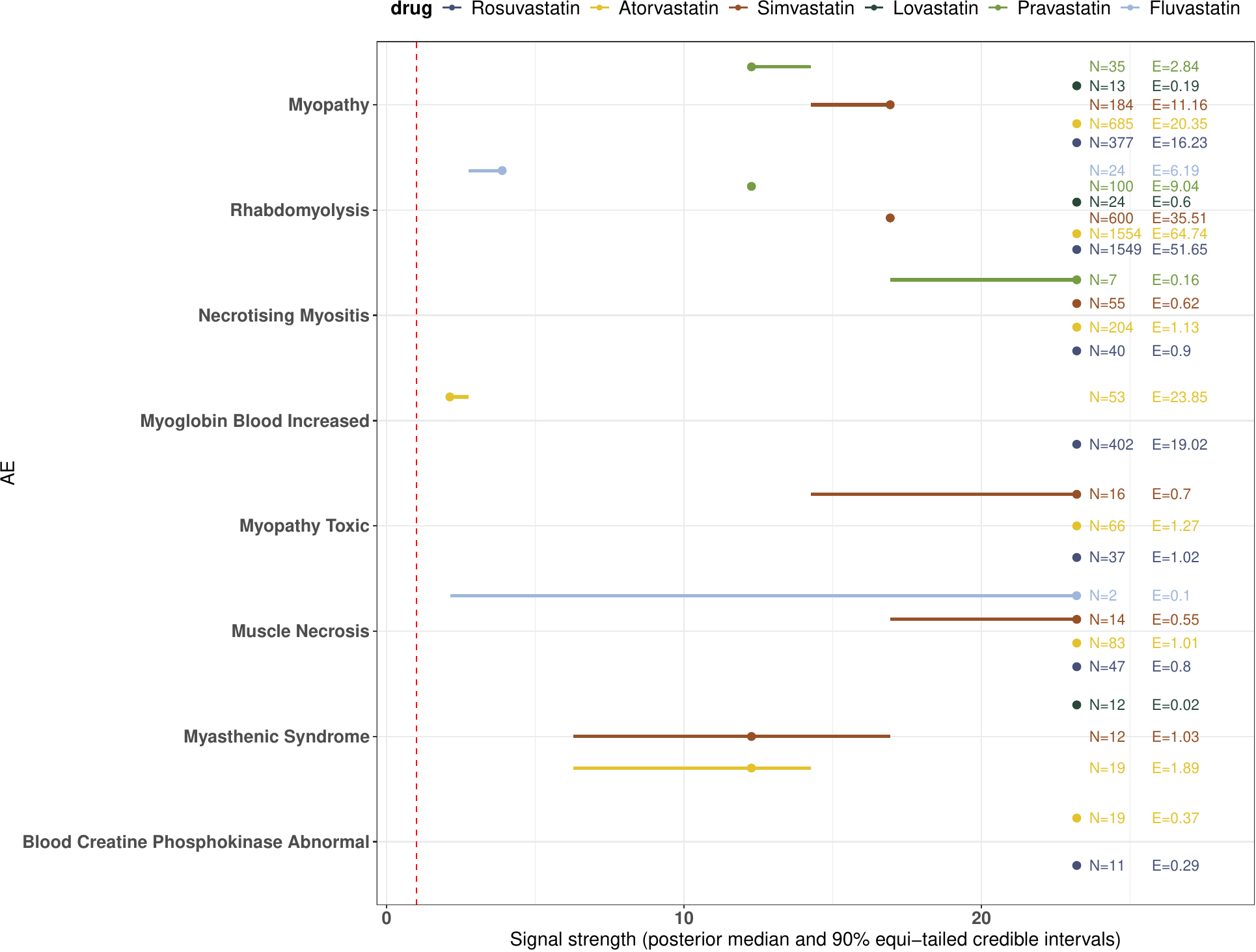} \end{center}

The above eyeplot visualizes empirical posterior inferences on \(10\) prominent AEs across \(6\) statin drugs through computed empirical Bayesian posterior distributions of signal strengths \(\{\lambda_{ij}\}\) obtained from the general-gamma model fitted on the statin2025\_44 dataset. The points and bars represent the posterior medians and \(90\%\) equi-tailed credible intervals for the corresponding AE-drug pair-specific \(\{\lambda_{ij}\}\), with different colors indicating the results from different statin drugs. The red dotted vertical line represents the value ``1'\,'. The texts on the right provide the number of observations as well as the null baseline expected counts under independence for an AE-drug pair. We note that, due to the shrinkage effect induced by the Dirichlet hyperprior, extremely large and sparse estimates of \(\lambda_{ij}\) are shrunk toward more moderate values. For example, the AE-drug pair ``Myasthenic Syndrome - Lovastatin,'' which has a raw \(O/E = 12/0.02 = 600\), is shrunk to approximately 25 under the general-gamma model.

Now, we increase the N\_threshold argument to 20 to focus on AE-drug combinations with stronger evidence. The associated codes are provided below.

\begin{verbatim}
eyeplot_N20_gg_tune_statin44 <- plot(gg_tune_statin44,
  type = "eyeplot",
  num_top_AEs = 8,
  N_threshold = 20,
  log_scale = FALSE,
  text_shift = 2.1,
  text_size = 4,
  x_lim_scalar = 1.2
) +
  guides(color = guide_legend(nrow = 1)) +
  theme(
    axis.text = element_text(size = 13, face = "bold"),
    legend.title = element_text(size = 14, face = "bold"),
    strip.text = element_text(size = 14, face = "bold"),
    axis.title.x = element_text(size = 14),
    axis.title.y = element_text(size = 14),
    legend.text = element_text(size = 14),
    legend.position = "top"
  )

eyeplot_N20_gg_tune_statin44
\end{verbatim}

\begin{center}\includegraphics[width=1\linewidth]{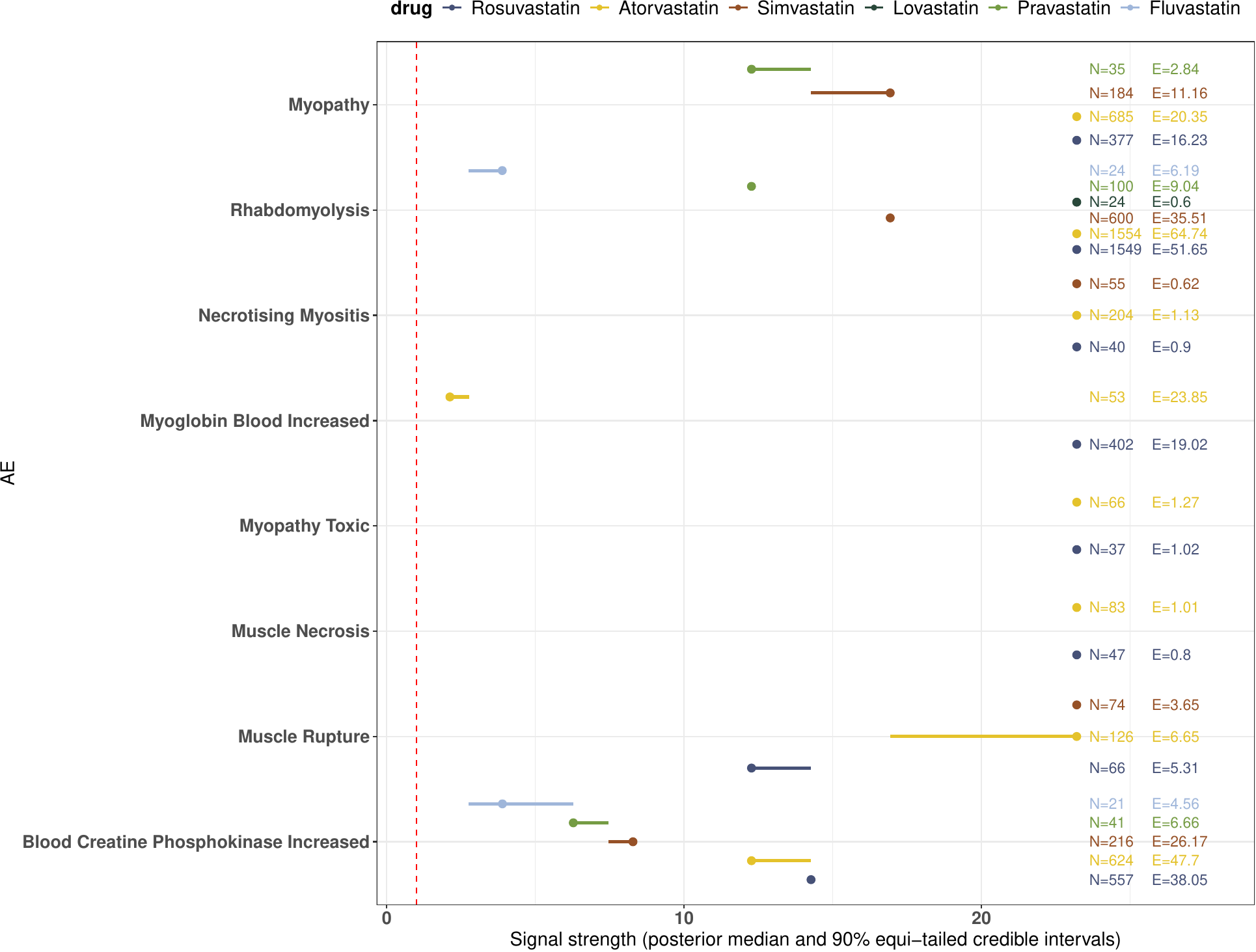} \end{center}

The following observations are made from the above plots. First, ``Rhabdomyolysis'', ``Myopathy'' and ``Necrotising Myositis'' are the top three most significant AEs. Rhabdomyolysis appears to have a significant association across all 6 statin drugs in both signal detection and estimation. Second, all of the top AEs are related to muscle which is consistent with the literature \citep{ramkumar2016statin}.

\paragraph{Analysis based on the KM model}\label{analysis-based-on-the-km-model}

We now consider fitting a KM model to identify and estimate signals in the statin2025\_44 data. To achieve this, the same pvEBayes() function can be used, with two changes in the input arguments: (1) model is now set to ``KM''; (2) argument alpha is not needed, and thus, there is no hyperparameter selection process, compared to general-gamma model. 

\begin{verbatim}
KM_statin44 <- pvEBayes(   
  statin2025_44,  
  model = "KM",
  n_posterior_draws = 10000
) 
\end{verbatim}

Similarly, print and summary methods can be applied to the fitted km model. The computation shows that 102 AE-drug combinations are detected as a signal.

\begin{verbatim}
KM_statin44

KM_statin44_detected_signal <- summary(KM_statin44,
  return = "detected signal"
)
sum(KM_statin44_detected_signal)
\end{verbatim}

\begin{verbatim}
#> [1] 102
\end{verbatim}

To visualize the signal detection and estimation result of the GPS model, we generate the heatmap and eyeplot using the following codes:

\begin{verbatim}
plot(KM_statin44,
  type = "heatmap",
  num_top_AEs = 10,
  cutoff = 1.001
) +
  theme(
    axis.text = element_text(size = 13, face = "bold"),
    legend.title = element_text(size = 14, face = "bold"),
    strip.text = element_text(size = 14, face = "bold"),
    legend.position = "top"
  )
\end{verbatim}

\begin{center}\includegraphics[width=1\linewidth]{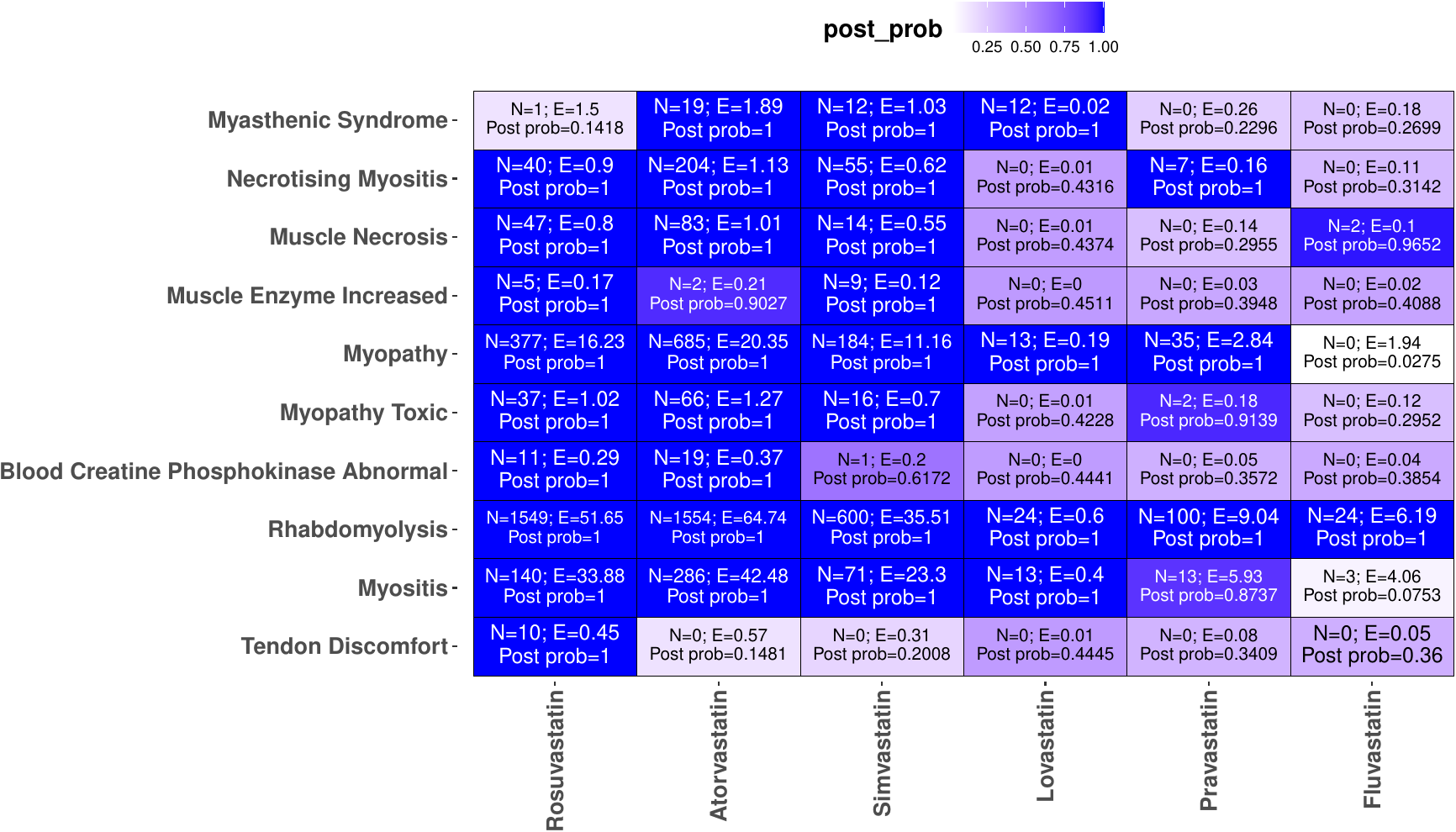} \end{center}

Unlike the general-gamma model, the KM model lacks a shrinkage mechanism to mitigate extremely large estimates of \(\lambda_{ij}\) that can arise from low expected null values---for example, when N = 2 and E = 0.01. We recommend setting a moderate N\_threshold argument, eg. 20, and plotting the posterior distribution of \(\lambda_{ij}\) in the natural logarithm scale. The relevant code is provided in the following.

\begin{verbatim}
plot(KM_statin44,
  type = "eyeplot",
  num_top_AEs = 8,
  N_threshold = 20,
  log_scale = TRUE,
  text_shift = 0.7,
  text_size = 4,
  x_lim_scalar = 1.2
) +
  guides(color = guide_legend(nrow = 2)) +
  theme(
    axis.text = element_text(size = 13, face = "bold"),
    legend.title = element_text(size = 14, face = "bold"),
    strip.text = element_text(size = 14, face = "bold"),
    axis.title.x = element_text(size = 14),
    axis.title.y = element_text(size = 14),
    legend.text = element_text(size = 14),
    legend.position = "top"
  )
\end{verbatim}

\begin{center}\includegraphics[width=1\linewidth]{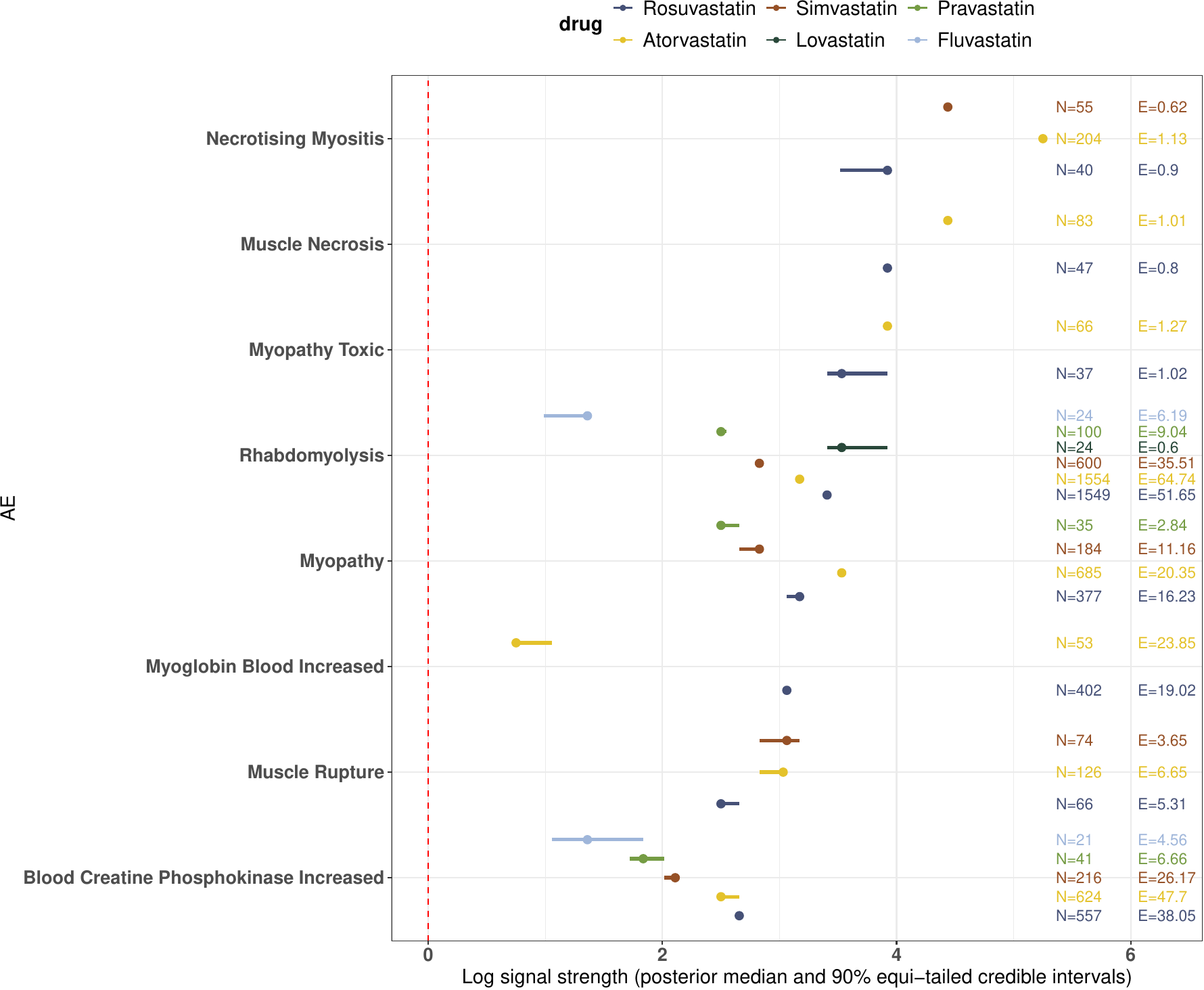} \end{center}

The following observations are made from the above plots. The top AEs that appeared in the heatmap and eyeplot are similar to that of general-gamma. The posterior estimation of \(\lambda_{ij}\) is close to raw \(O/E\).

\paragraph{Analysis based on the Efron model}\label{analysis-based-on-the-efron-model}

We are going to fit an Efron model to identify and estimate signals in the statin2025\_44 data. Similar to the general-gamma model, the Efron model also has hyperparameters \((p, c_0)\) to be tuned. To achieve this, pvEBayes\_tune() function can be used.

\begin{verbatim}
E_tune_statin44 <- pvEBayes_tune(statin2025_44,
  model = "efron",
  p_vec = c(40, 60, 80, 100, 120),
  c0_vec = c(1e-5, 1e-4, 1e-3, 1e-2, 1e-1),
  use_AIC = TRUE,
  n_posterior_draws = 10000
)
\end{verbatim}

\begin{verbatim}
E_tune_statin44

E_tune_statin44_detected_signal <- summary(E_tune_statin44,
  return = "detected signal"
)
sum(E_tune_statin44_detected_signal)
\end{verbatim}

\begin{verbatim}
#> [1] 102
\end{verbatim}

Similarly, print and summary methods can be applied to the fitted km model. The computation shows that 102 AE-drug combinations are detected as a signal.

\begin{verbatim}
plot(E_tune_statin44,
  type = "heatmap",
  num_top_AEs = 10,
  cutoff = 1.001
) +
  theme(
    axis.text = element_text(size = 13, face = "bold"),
    legend.title = element_text(size = 14, face = "bold"),
    strip.text = element_text(size = 14, face = "bold"),
    legend.position = "top"
  )
\end{verbatim}

\begin{center}\includegraphics[width=1\linewidth]{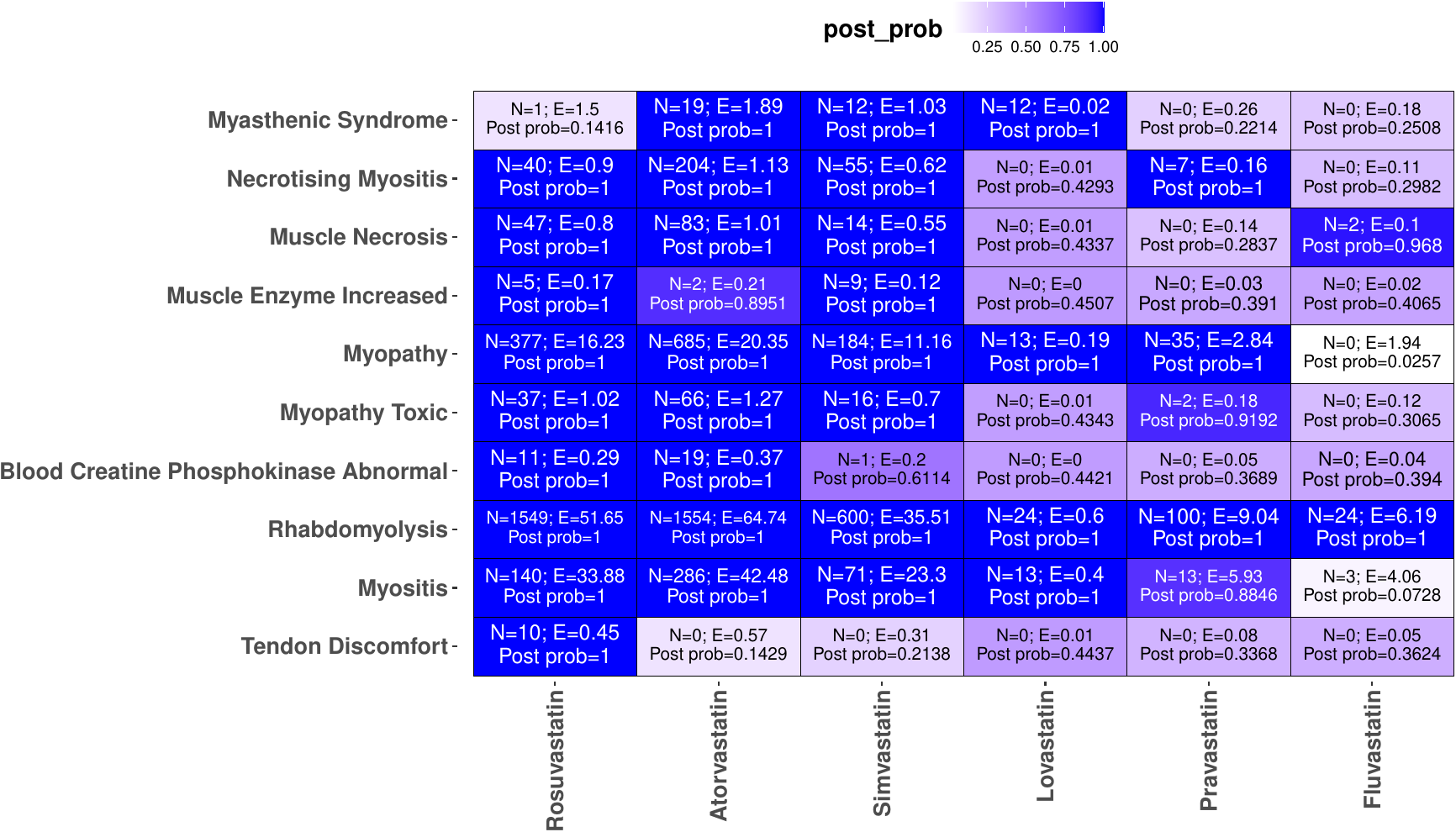} \end{center}

Similar to the KM model, we suggest a moderate N\_threshold setting.

\begin{verbatim}
plot(E_tune_statin44,
  type = "eyeplot",
  num_top_AEs = 8,
  N_threshold = 20,
  log_scale = TRUE,
  text_shift = 0.7,
  text_size = 4,
  x_lim_scalar = 1.2
) +
  guides(color = guide_legend(nrow = 2)) +
  theme(
    axis.text = element_text(size = 13, face = "bold"),
    legend.title = element_text(size = 14, face = "bold"),
    strip.text = element_text(size = 14, face = "bold"),
    axis.title.x = element_text(size = 14),
    axis.title.y = element_text(size = 14),
    legend.text = element_text(size = 14),
    legend.position = "top"
  )
\end{verbatim}

\begin{center}\includegraphics[width=1\linewidth]{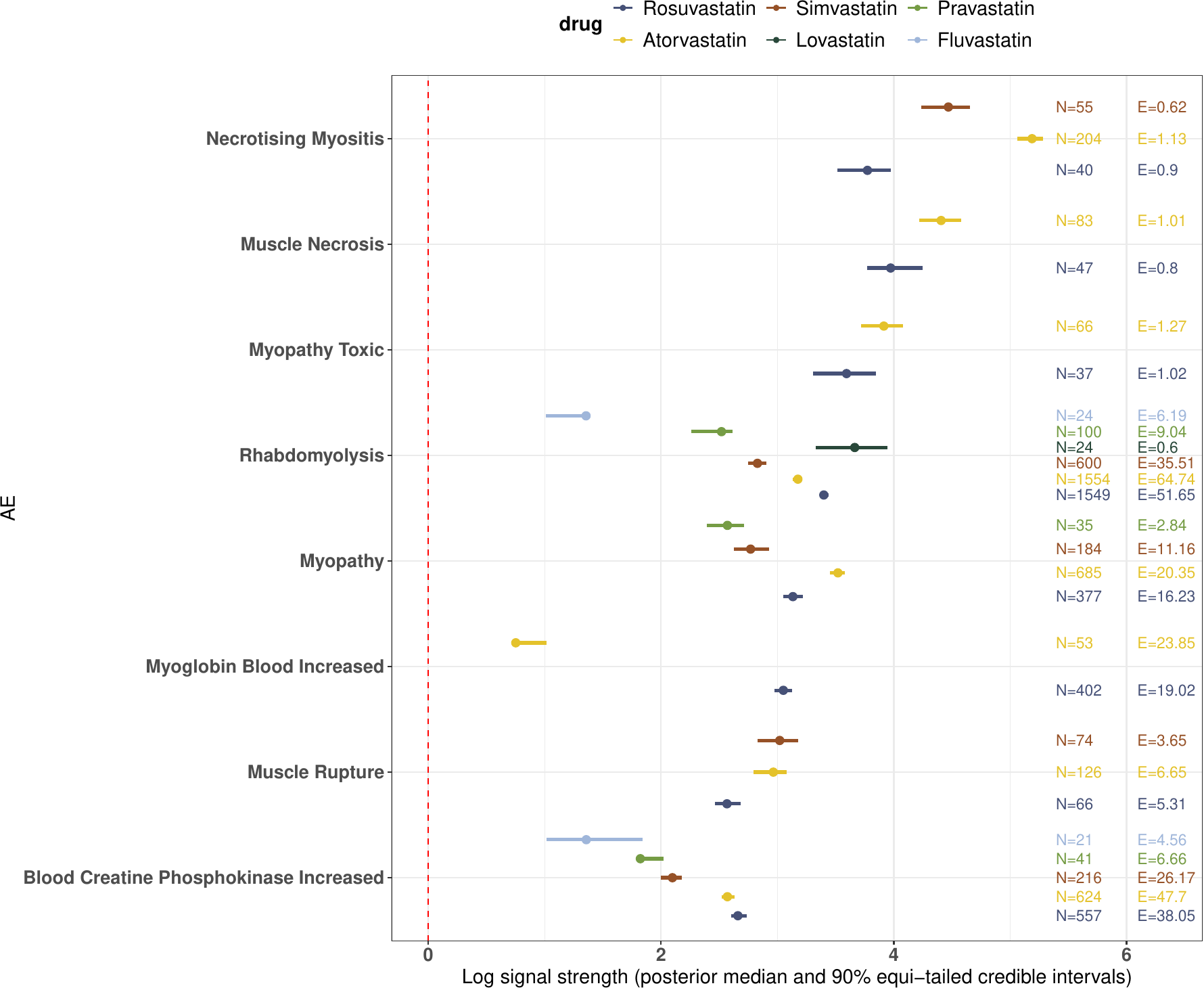} \end{center}

The heatmap and eyeplot visualization for the Efron model is very similar to that for the KM model. The shrinkage of the Efron model makes the estimated prior approach to the uniform distribution. As a result, the credible intervals shown in the eyeplot are wider than the KM model.

\paragraph{Analysis based on GPS(2-gamma) model}\label{analysis-based-on-gps2-gamma-model}

We now consider fitting a GPS(2-gamma) model to identify and estimate signals in the statin2025\_44 data. To achieve this, the same pvEBayes() function can be used, with two changes in the input arguments: (1) model is now set to ``GPS''; (2) argument alpha is not needed, and thus, there is no hyperparameter selection process, compared to general-gamma model. The following R codes are used:

\begin{verbatim}
gps_statin44 <- pvEBayes(
  statin2025_44,
  model = "GPS",
  n_posterior_draws = 10000
)
\end{verbatim}

Similarly, print and summary methods can be applied to the fitted GPS model. The computation shows that 88 AE-drug combinations are detected as a signal.

\begin{verbatim}
gps_statin44

gps_statin44_detected_signal <- summary(gps_statin44,
  return = "detected signal"
)
sum(gps_statin44_detected_signal)
\end{verbatim}

\begin{verbatim}
#> [1] 87
\end{verbatim}

To visualize the signal detection and estimation result of the GPS model, we generate the heatmap and eyeplot using the following codes:

\begin{verbatim}
plot(gps_statin44,
  type = "heatmap",
  num_top_AEs = 10,
  cutoff = 1.001
) +
  theme(
    axis.text = element_text(size = 13, face = "bold"),
    legend.title = element_text(size = 14, face = "bold"),
    strip.text = element_text(size = 14, face = "bold"),
    legend.position = "top"
  )
\end{verbatim}

\begin{center}\includegraphics[width=1\linewidth]{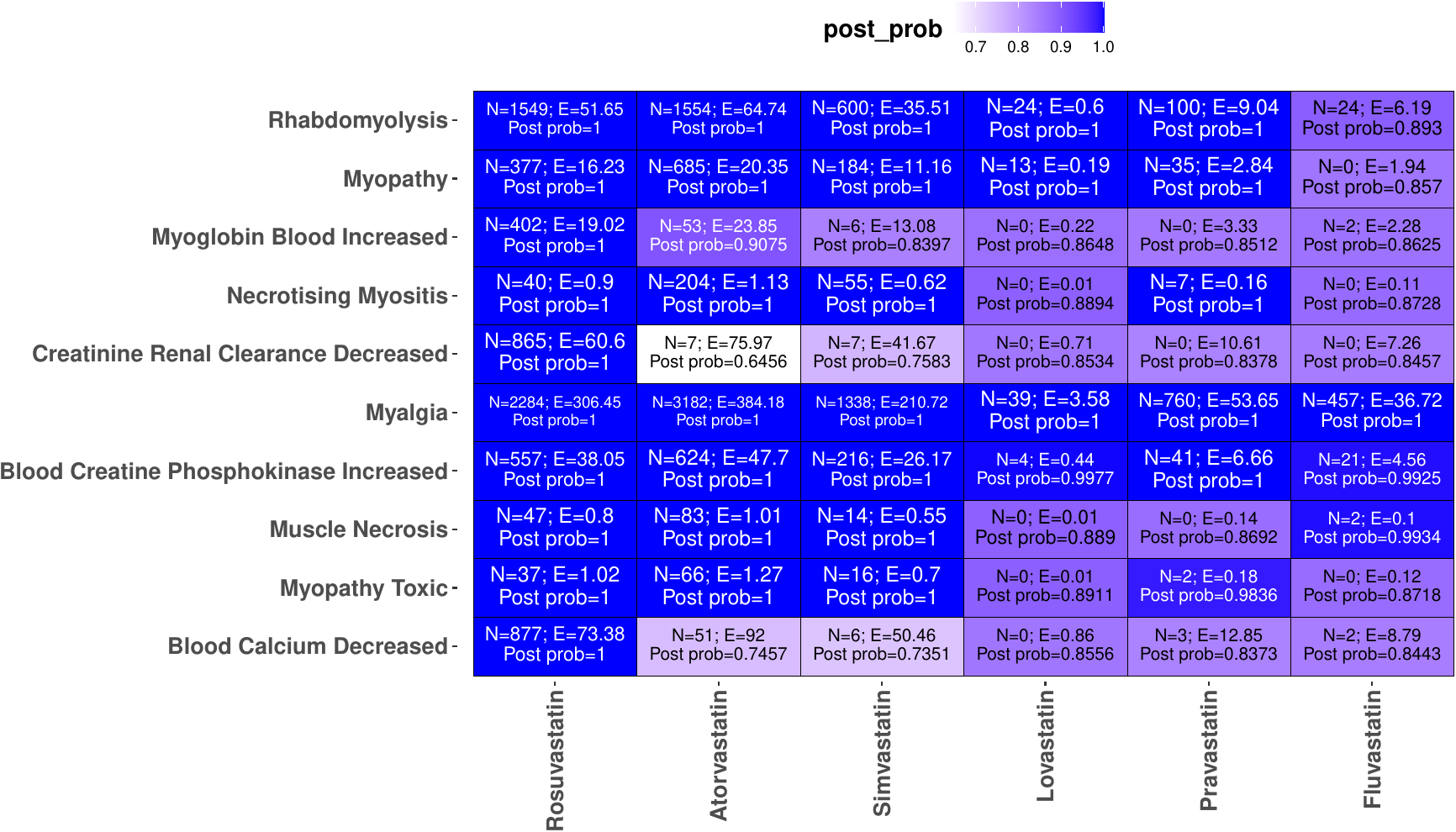} \end{center}

\begin{verbatim}
plot(gps_statin44,
  type = "eyeplot",
  num_top_AEs = 8,
  N_threshold = 20,
  log_scale = FALSE,
  text_shift = 1,
  text_size = 4,
  x_lim_scalar = 1.15
) +
  guides(color = guide_legend(nrow = 2)) +
  theme(
    axis.text = element_text(size = 13, face = "bold"),
    legend.title = element_text(size = 14, face = "bold"),
    strip.text = element_text(size = 14, face = "bold"),
    axis.title.x = element_text(size = 14),
    axis.title.y = element_text(size = 14),
    legend.text = element_text(size = 14),
    legend.position = "top"
  )
\end{verbatim}

\begin{center}\includegraphics[width=1\linewidth]{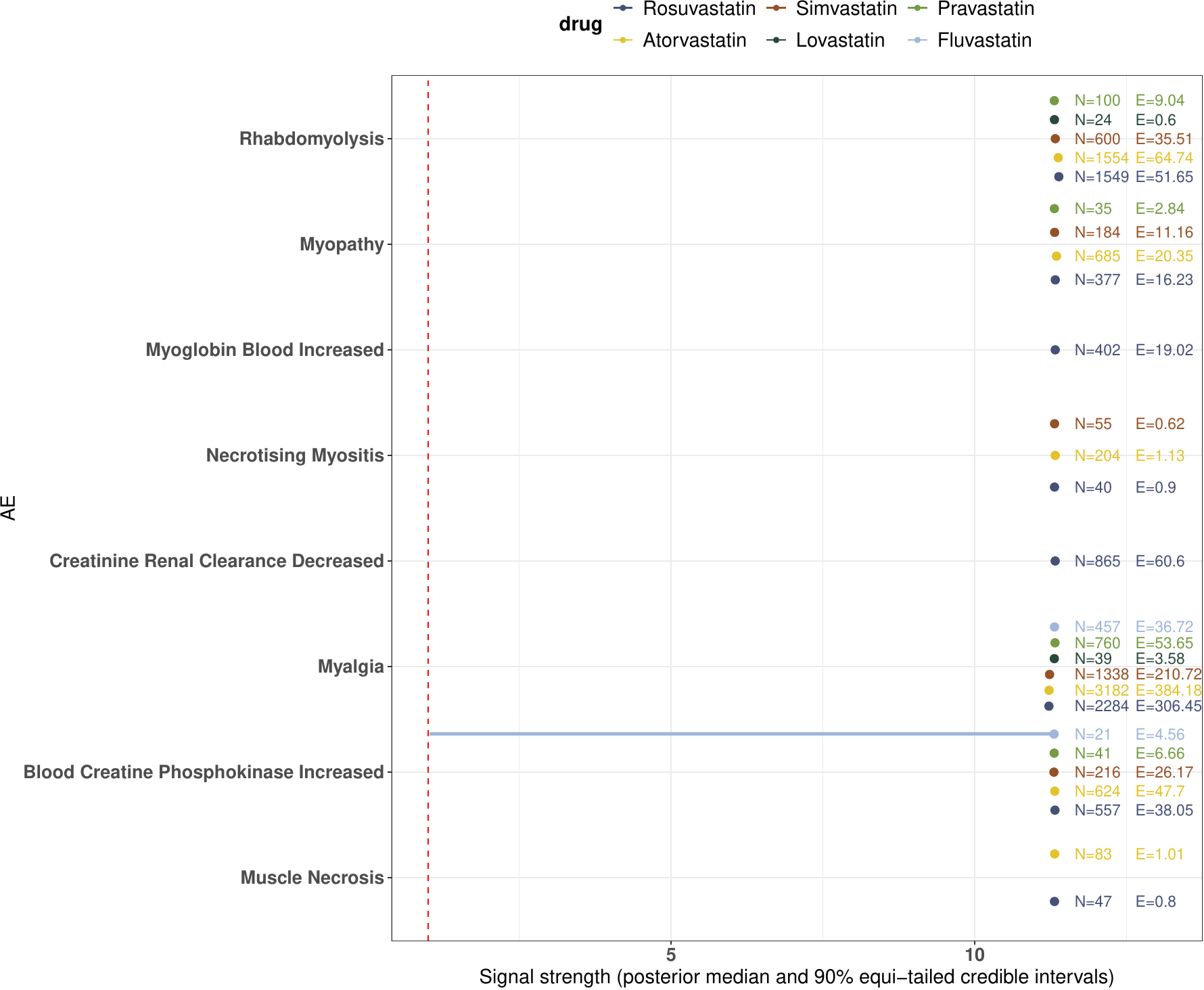} \end{center}

The following observations are made: (1) The GPS model detects fewer signals compared to the general-gamma model. ``Myalgia'' and ``Blood Creatine Phosphokinase Increased'' are the only adverse events identified as significant across all six statin drugs. Overall, the signal detection results of the GPS model are similar to those of the general-gamma model. (2) The GPS model employs a two-component gamma mixture prior, which may be sufficient for signal detection. However, its estimation of signal strength shows very wide credible intervals, and the posterior distributions are restricted near 11, implying limited flexibility in capturing the structure of SRS data.

\subsection{Analysis of the gbca2025\_69 data set}\label{analysis-of-the-gbca2025_69-data-set}

We next analyze the gbca2025\_69 dataset with 70 AE rows and 8 drug columns. This data set has 256 zeros and 408 entries have observations less or equal to 5 among 560 total entries. This implies zero-inflation and sparse signals in the dataset, which provide a justification for using the general-gamma model. The use of the general-gamma model begins with hyperparameter \(\alpha\) selection. The codes are provided in the following.

\begin{verbatim}
gg_gbca <- pvEBayes_tune(gbca2025_69,
  use_AIC = TRUE
)
\end{verbatim}

\begin{verbatim}
#> The alpha value selected under AIC is 0.5,
#> The alpha value selected under BIC is 0.1.
\end{verbatim}

\begin{verbatim}
#>   alpha      AIC      BIC num_mixture
#> 1   0.0 2711.382 2802.269           7
#> 2   0.1 2711.232 2802.118           7
#> 3   0.3 2725.211 2816.098           7
#> 4   0.5 2698.868 2841.690          11
#> 5   0.7 2710.707 2879.497          13
#> 6   0.9 2836.471 3277.921          34
\end{verbatim}

Again, we use \(\alpha = 0.1\) as suggested by BIC and fit the general-gamma model.

\begin{verbatim}
gg_gbca
\end{verbatim}

\begin{verbatim}
gg_gbca_detected_signal <- summary(gg_gbca,
  return = "detected signal"
)
sum(gg_gbca_detected_signal)
\end{verbatim}

\begin{verbatim}
#> [1] 57
\end{verbatim}

To get a visual understanding of the results we generate the heatmap and eyeplot in the following.

\begin{verbatim}
plot(gg_gbca,
  type = "heatmap",
  num_top_AEs = 10,
  num_top_drugs = 5,
  cutoff = 1.001
)
\end{verbatim}

\begin{center}\includegraphics[width=1\linewidth]{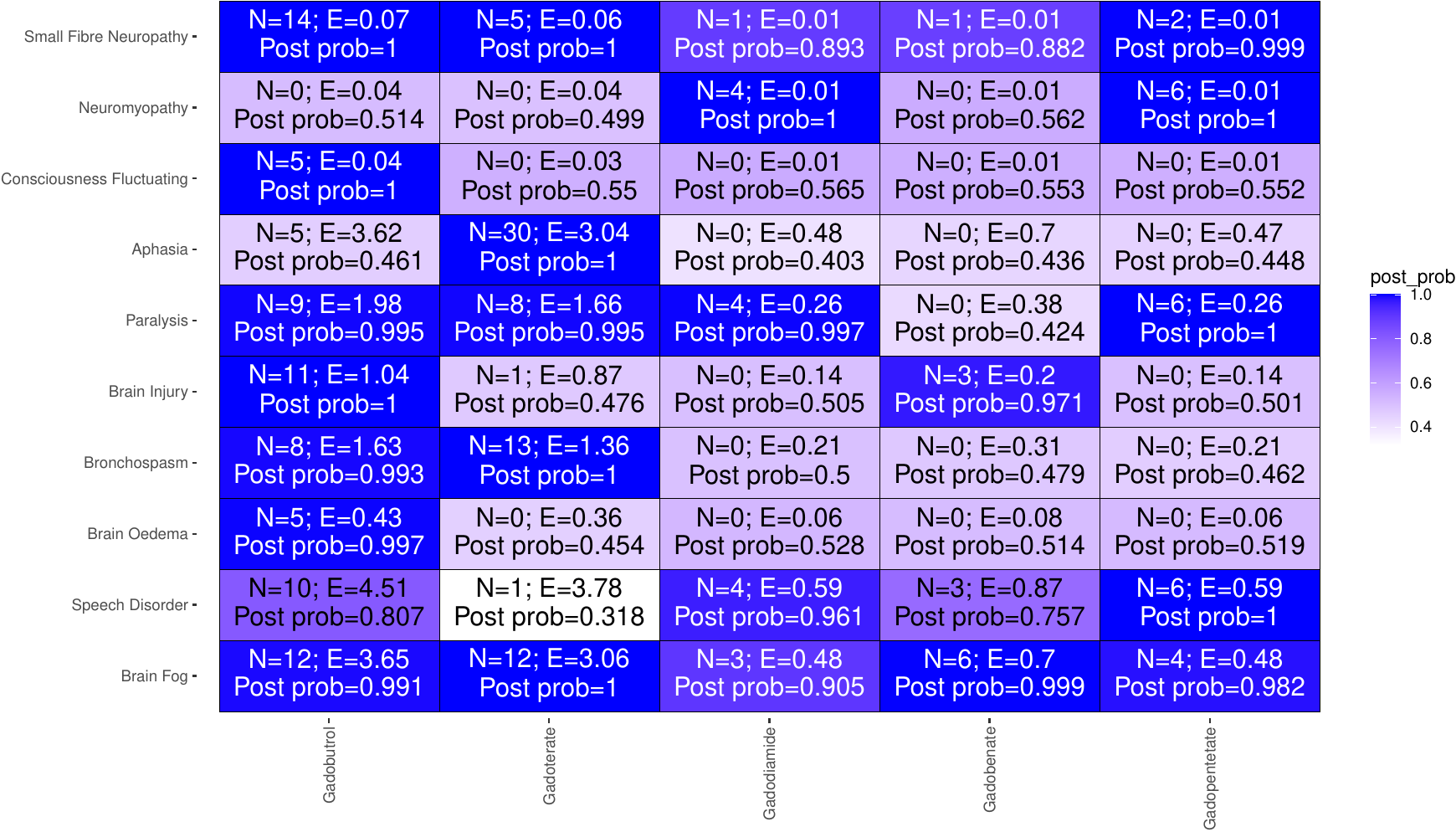} \end{center}

Due to the sparsity of the gbca69 dataset, we set the argument N\_threshold to 10 to have enough AE presented in the eyeplot.

\begin{verbatim}
eyeplot_gg_tune_gbca <- plot(gg_gbca,
  type = "eyeplot",
  num_top_AEs = 8,
  N_threshold = 10,
  log_scale = FALSE,
  text_shift = 5,
  text_size = 4,
  x_lim_scalar = 1.1
) +
  theme(
    axis.text = element_text(size = 13, face = "bold"),
    legend.title = element_text(size = 14, face = "bold"),
    strip.text = element_text(size = 14, face = "bold"),
    axis.title.x = element_text(size = 14),
    axis.title.y = element_text(size = 14),
    legend.text = element_text(size = 14),
    legend.position = "top"
  )

eyeplot_gg_tune_gbca
\end{verbatim}

\begin{center}\includegraphics[width=1\linewidth]{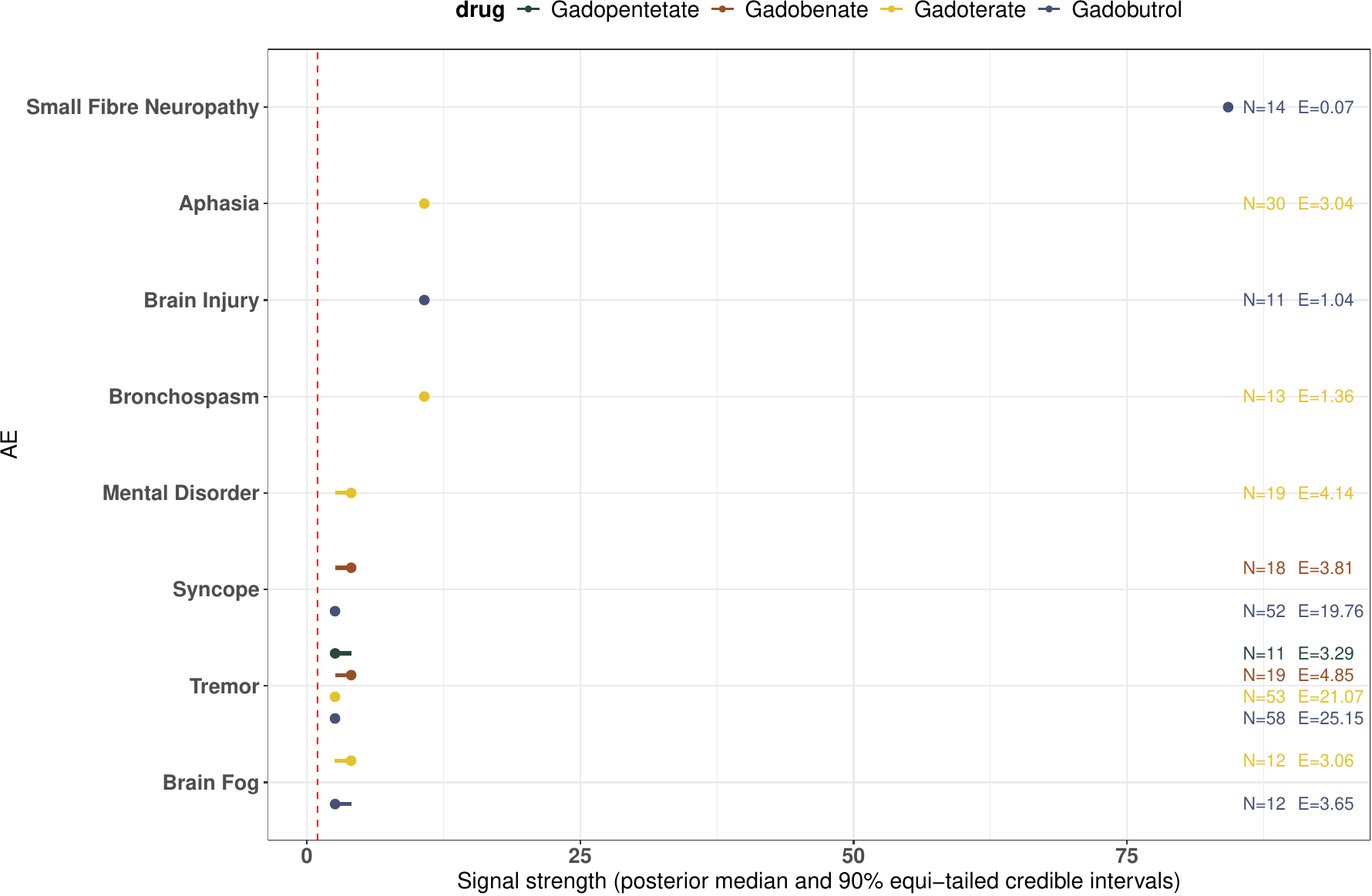} \end{center}

The following observations are made. The low observations in the gbca2025\_69 dataset imply that the signals in gbca69 are sparse. Consequently, most of the AE-drug pairs in the top 10 AEs have relatively low estimated \(\lambda_{ij}\), compared with the statin dataset. In the visualization of the results we still found some prominent signals observed in (Small Fibre Neuropathy, Gadopentetate), (Aphasia, Gadoterate), (Brain Injury, Gadopentetate) and (Bronchospasm, Gadoterate).

\section{Discussion}\label{discussion}

As extensions of the GPS model, nonparametric empirical Bayes models have been shown to perform well in both detecting and estimating signals in drug safety data \citep{tan2025flexibleempiricalbayesianapproaches}. The lack of comprehensive software implementation of these models has limited its applicability in practice.

The package \CRANpkg{pvEBayes} is our contribution to this area, providing an easily accessible platform that implements GPS, K-gamma, general-gamma, KM, and Efron models. In addition, it includes tools for random table generation and result visualization.

\section{Acknowledgment}\label{acknowledgment}

We thank Raktim Mukhopadhyay for his valuable suggestions on the R package development, particularly for package testing and debugging.

Marianthi Markatou acknowledges funding from Kaleida Health Foundation (award number 82114) that partially supported the work of the first author.

\bibliographystyle{abbrv}
\bibliography{references}  

@Manual{pvEBayes_pkg,
  title = {pvEBayes: Empirical Bayes Methods for Pharmacovigilance},
  author = {Yihao Tan and Marianthi Markatou and Saptarshi Chakraborty and Raktim Mukhopadhyay},
  year = {2025},
  note = {R package version 0.1.1},
  url = {https://CRAN.R-project.org/package=pvEBayes},
  doi = {10.32614/CRAN.package.pvEBayes},
}

@article{amery1999there,
  title={Why there is a need for pharmacovigilance},
  author={Amery, Willem K},
  journal={Pharmacoepidemiology and Drug Safety},
  volume={8},
  number={1},
  pages={61--64},
  year={1999},
  publisher={Wiley Online Library}
}

@article{markatou2014pattern,
  title={A pattern discovery framework for adverse event evaluation and inference in spontaneous reporting systems},
  author={Markatou, Marianthi and Ball, Robert},
  journal={Statistical Analysis and Data Mining: The ASA Data Science Journal},
  volume={7},
  number={5},
  pages={352--367},
  year={2014},
  publisher={Wiley Online Library}
}

@article{bate1998bayesian,
  title={A {Bayesian} neural network method for adverse drug reaction signal generation},
  author={Bate, Andrew and Lindquist, Marie and Edwards, I Ralph and Olsson, Sten and Orre, Roland and Lansner, Anders and De Freitas, R Melhado},
  journal={European Journal of Clinical Pharmacology},
  volume={54},
  pages={315--321},
  year={1998},
  publisher={Springer}
}

@article{ding2020evaluation,
  title={An evaluation of statistical approaches to postmarketing surveillance},
  author={Ding, Yuxin and Markatou, Marianthi and Ball, Robert},
  journal={Statistics in Medicine},
  volume={39},
  number={7},
  pages={845--874},
  year={2020},
  publisher={Wiley Online Library}
}

@article{huang2011likelihood,
  title={A likelihood ratio test based method for signal detection with application to {FDA}{\textquoteright}s drug safety data},
  author={Huang, Lan and Zalkikar, Jyoti and Tiwari, Ram C},
  journal={Journal of the American Statistical Association},
  volume={106},
  number={496},
  pages={1230--1241},
  year={2011},
  publisher={Taylor \& Francis}
}

@article{hu2015signal,
  title={Signal detection in {FDA} {AERS} database using {Dirichlet} process},
  author={Hu, Na and Huang, Lan and Tiwari, Ram C},
  journal={Statistics in Medicine},
  volume={34},
  number={19},
  pages={2725--2742},
  year={2015},
  publisher={Wiley Online Library}
}

@article{zhao2018extended,
  title={Extended likelihood ratio test-based methods for signal detection in a drug class with application to {FDA}{\textquoteright}s adverse event reporting system database},
  author={Zhao, Yueqin and Yi, Min and Tiwari, Ram C},
  journal={Statistical Methods in Medical Research},
  volume={27},
  number={3},
  pages={876--890},
  year={2018},
  publisher={SAGE Publications Sage UK: London, England}
}

@article{huang2017zero,
  title={Zero-inflated {Poisson} model based likelihood ratio test for drug safety signal detection},
  author={Huang, Lan and Zheng, Dan and Zalkikar, Jyoti and Tiwari, Ram},
  journal={Statistical Methods in Medical Research},
  volume={26},
  number={1},
  pages={471--488},
  year={2017},
  publisher={SAGE Publications Sage UK: London, England}
}

@article{evans2001use,
  title={Use of proportional reporting ratios ({PRRs}) for signal generation from spontaneous adverse drug reaction reports},
  author={Evans, Stephen JW and Waller, Patrick C and Davis, S},
  journal={Pharmacoepidemiology and Drug Safety},
  volume={10},
  number={6},
  pages={483--486},
  year={2001},
  publisher={Wiley Online Library}
}

@article{rothman2004reporting,
  title={The reporting odds ratio and its advantages over the proportional reporting ratio},
  author={Rothman, Kenneth J and Lanes, Stephan and Sacks, Susan T},
  journal={Pharmacoepidemiology and Drug Safety},
  volume={13},
  number={8},
  pages={519--523},
  year={2004},
  publisher={Wiley Online Library}
}

@article{huang2013likelihood,
  title={Likelihood ratio test-based method for signal detection in drug classes using {FDA}{\textquoteright}s {AERS} database},
  author={Huang, Lan and Zalkikar, Jyoti and Tiwari, Ram C},
  journal={Journal of Biopharmaceutical Statistics},
  volume={23},
  number={1},
  pages={178--200},
  year={2013},
  publisher={Taylor \& Francis}
}

@article{chakraborty2022use,
  title={On the use of the likelihood ratio test methodology in pharmacovigilance},
  author={Chakraborty, Saptarshi and Liu, Anran and Ball, Robert and Markatou, Marianthi},
  journal={Statistics in Medicine},
  volume={41},
  number={27},
  pages={5395--5420},
  year={2022},
  publisher={Wiley Online Library}
}

@article{dumouchel1999bayesian,
  title={{Bayesian} data mining in large frequency tables, with an application to the {FDA} spontaneous reporting system},
  author={DuMouchel, William},
  journal={The American Statistician},
  volume={53},
  number={3},
  pages={177--190},
  year={1999},
  publisher={Taylor \& Francis}
}

@article{tan2025flexibleempiricalbayesianapproaches,
author = {Tan, Yihao and Markatou, Marianthi and Chakraborty, Saptarshi},
title = {Flexible Empirical Bayesian Approaches to Pharmacovigilance for Simultaneous Signal Detection and Signal Strength Estimation in Spontaneous Reporting Systems Data},
journal = {Statistics in Medicine},
volume = {44},
number = {18-19},
pages = {e70195},
doi = {10.1002/sim.70195},
url = {https://onlinelibrary.wiley.com/doi/abs/10.1002/sim.70195},
eprint = {https://onlinelibrary.wiley.com/doi/pdf/10.1002/sim.70195},
year = {2025}

}

@article{amrheinScientistsRiseStatistical2019,
	title = {Scientists rise up against statistical significance},
	volume = {567},
	copyright = {2021 Nature},
	url = {https://www.nature.com/articles/d41586-019-00857-9},
	doi = {10.1038/d41586-019-00857-9},
	language = {en},
	number = {7748},
	urldate = {2024-08-13},
	journal = {Nature},
	author = {Amrhein, Valentin and Greenland, Sander and McShane, Blake},
	month = mar,
	year = {2019},
	pages = {305--307},
}

@article{gelmanStatisticalCrisisScience2016,
	title = {The {Statistical} {Crisis} in {Science}},
	isbn = {978-1-4008-7337-1},
	urldate = {2024-08-13},
	journal = {The {Best} {Writing} on {Mathematics}},
	publisher = {Princeton University Press},
	author = {Gelman, Andrew and Loken, Eric},
	month = dec,
	year = {2016},
	doi = {10.1515/9781400873371-028},
	pages = {305--318}
}

@article{rothmanDisengagingStatisticalSignificance2016,
	title = {Disengaging from statistical significance},
	volume = {31},
	issn = {1573-7284},
	url = {https://doi.org/10.1007/s10654-016-0158-2},
	doi = {10.1007/s10654-016-0158-2},
	language = {en},
	number = {5},
	urldate = {2024-08-13},
	journal = {European Journal of Epidemiology},
	author = {Rothman, Kenneth J.},
	month = may,
	year = {2016},
	pages = {443--444}
}

@article{wassersteinASAStatementPValues2016,
	title = {The {ASA} {Statement} on p-{Values}: {Context}, {Process}, and {Purpose}},
	volume = {70},
	issn = {0003-1305},
	shorttitle = {The {ASA} {Statement} on p-{Values}},
	url = {https://doi.org/10.1080/00031305.2016.1154108},
	doi = {10.1080/00031305.2016.1154108},
	number = {2},
	urldate = {2024-08-13},
	journal = {The American Statistician},
	author = {Wasserstein, Ronald L. and Lazar, Nicole A.},
	month = apr,
	year = {2016},
	pages = {129--133}
}

@article{liu2024mddcrpythonpackage,
      title={{MDDC}: An {R} and {Python} Package for Adverse Event Identification in Pharmacovigilance Data},
      author={Anran Liu and Raktim Mukhopadhyay and Marianthi Markatou},
      year={2025},
      journal={Scientific Reports},
      volume    = {15},
      number    = {1},
      pages     = {21317},
      doi       = {10.1038/s41598-025-00635-w}

}

@article{li2020adverseeventenrichmenttests,
      title={Vaccine adverse event enrichment tests},
      author={Shuoran Li and Lili Zhao},
      year={2021},
      journal={Statistics in Medicine},
      volume={40},
      number={19},
      pages={4269-4278},
      doi = {https://doi.org/10.1002/sim.9027}

}

@article{shih2010sequential,
  title={Sequential generalized likelihood ratio tests for vaccine safety evaluation},
  author={Shih, Mei-Chiung and Lai, Tze Leung and Heyse, Joseph F and Chen, Jie},
  journal={Statistics in Medicine},
  volume={29},
  number={26},
  pages={2698--2708},
  year={2010},
  publisher={Wiley Online Library}
}

@article{courtois2021new,
  title={New adaptive lasso approaches for variable selection in automated pharmacovigilance signal detection},
  author={Courtois, {\'E}meline and Tubert-Bitter, Pascale and Ahmed, Isma{\"\i}l},
  journal={BMC Medical Research Methodology},
  volume={21},
  pages={1--17},
  year={2021},
  publisher={Springer}
}

@Manual{pkg_phvid,
    title = {{PhViD: an R package for PharmacoVigilance signal Detection}},
    author = {I. Ahmed and A. Poncet},
    year = {2016},
    note = {R package version 1.0.8},
}

@Article{canida2017,
    author = {Travis Canida and John Ihrie},
    title = {{openEBGM: An R Implementation of the Gamma-Poisson Shrinker Data Mining Model}},
    year = {2017},
    journal = {The R Journal},
    volume = {9},
    number = {2},
    pages = {499-519},
    url = {https://journal.r-project.org/archive/2017/RJ-2017-063/index.html},
  }

@Manual{pkg_pvLRT,
    title = {{pvLRT}: {Likelihood} Ratio Test-Based Approaches to Pharmacovigilance},
    author = {Saptarshi Chakraborty and Marianthi Markatou},
    year = {2022},
    note = {R package version 0.4},
}

@Manual{pkg_sglr,
    title = {{sglr:   An R package for computing the boundaries for sequential generalized
  likelihood ratio test for pre-licensure vaccine studies}},
    author = {Balasubramanian Narasimhan and Mei-Chiung Shih},
    year = {2012},
    note = {R package version 0.7},
    url = {http://CRAN.R-project.org/package=sglr},
}

@Manual{pkg_sequential,
    title = {{Sequential: Exact Sequential Analysis for Poisson and Binomial Data}},
    author = {Ivair Ramos Silva and Martin Kulldorff},
    year = {2021},
    note = {R package version 4.1},
    url = {https://CRAN.R-project.org/package=Sequential},
  }

@Manual{pkg_aeenrich,
    title = {{AEenrich: Adverse Event Enrichment Tests}},
    author = {Shuoran Li and Hongfan Chen and Lili Zhao and Michael Kleinsasser},
    year = {2021},
    note = {R package version 1.1.0},
    url = {https://CRAN.R-project.org/package=AEenrich},
}

@Manual{pkg_mds,
    title = {{mds: Medical Devices Surveillance}},
    author = {Gary Chung},
    year = {2020},
    note = {R package version 0.3.2},
    url = {https://CRAN.R-project.org/package=mds},
}

@Manual{Rmosek_manual,
  title = {Rmosek: The R-to-MOSEK optimization interface},
  author = {MOSEK ApS},
  year = {2025},
  note = {R package version 11.0.9},
  url = {http://www.mosek.com/},
}

@Manual{pkg_adapt4pv,
    title = {adapt4pv: Adaptive Approaches for Signal Detection in Pharmacovigilance},
    author = {Ismaïl Ahmed},
    year = {2023},
    note = {R package version 0.2-3},
    url = {https://CRAN.R-project.org/package=adapt4pv},
  }

@Article{pkg_KM,
    title = {{REBayes}: An {R} Package for Empirical {Bayes} Mixture Methods},
    author = {Roger Koenker and Jiaying Gu},
    journal = {Journal of Statistical Software},
    year = {2017},
    volume = {82},
    number = {8},
    pages = {1--26},
    doi = {10.18637/jss.v082.i08},
  }

@Manual{pkg_rcpp,
  title = {Rcpp: Seamless R and C++ Integration},
  author = {Dirk Eddelbuettel and Romain Francois and JJ Allaire and Kevin Ushey and Qiang Kou and Nathan Russell and Iñaki Ucar and Doug Bates and John Chambers},
  year = {2025},
  note = {R package version 1.0.14},
  url = {https://CRAN.R-project.org/package=Rcpp},
}

@Article{pkg_rcppeigen,
  title = {Fast and Elegant Numerical Linear Algebra Using the {RcppEigen} Package},
  author = {Douglas Bates and Dirk Eddelbuettel},
  journal = {Journal of Statistical Software},
  year = {2013},
  volume = {52},
  number = {5},
  pages = {1--24},
  doi = {10.18637/jss.v052.i05},
}

@Article{pkg_Efron,
    title = {{deconvolveR}: A $G$-Modeling Program for Deconvolution and Empirical {Bayes} Estimation},
    author = {Balasubramanian Narasimhan and Bradley Efron},
    journal = {Journal of Statistical Software},
    year = {2020},
    volume = {94},
    number = {11},
    pages = {1--20},
    doi = {10.18637/jss.v094.i11},
  }

@article{koenker2014convex,
  title={Convex optimization, shape constraints, compound decisions, and empirical {Bayes} rules},
  author={Koenker, Roger and Mizera, Ivan},
  journal={Journal of the American Statistical Association},
  volume={109},
  number={506},
  pages={674--685},
  year={2014},
  publisher={Taylor \& Francis}
}

@article{efron2016empirical,
  title={Empirical {Bayes} deconvolution estimates},
  author={Efron, Bradley},
  journal={Biometrika},
  volume={103},
  number={1},
  pages={1--20},
  year={2016},
  publisher={Oxford University Press}
}

@Book{pkg_ggplot2,
    author = {Hadley Wickham},
    title = {ggplot2: Elegant Graphics for Data Analysis},
    publisher = {Springer-Verlag New York},
    year = {2016},
    isbn = {978-3-319-24277-4},
    url = {https://ggplot2.tidyverse.org},
}

@article{efron2014two,
  title={Two modeling strategies for empirical {Bayes} estimation},
  author={Efron, Bradley},
  journal={{Statistical Science}: a review journal of the Institute of Mathematical Statistics},
  volume={29},
  number={2},
  pages={285},
  year={2014}
}

@article{fruhwirth2019here,
  title={From here to infinity: sparse finite versus {Dirichlet} process mixtures in model-based clustering},
  author={Fr{\"u}hwirth-Schnatter, Sylvia and Malsiner-Walli, Gertraud},
  journal={Advances in Data Analysis and Classification},
  volume={13},
  pages={33--64},
  year={2019},
  publisher={Springer}
}

@article{malsiner2016model,
  title={Model-based clustering based on sparse finite {Gaussian} mixtures},
  author={Malsiner-Walli, Gertraud and Fr{\"u}hwirth-Schnatter, Sylvia and Gr{\"u}n, Bettina},
  journal={Statistics and Computing},
  volume={26},
  number={1},
  pages={303--324},
  year={2016},
  publisher={Springer}
}

@article{malsiner2017identifying,
  title={Identifying mixtures of mixtures using {Bayesian} estimation},
  author={Malsiner-Walli, Gertraud and Fr{\"u}hwirth-Schnatter, Sylvia and Gr{\"u}n, Bettina},
  journal={Journal of Computational and Graphical Statistics},
  volume={26},
  number={2},
  pages={285--295},
  year={2017},
  publisher={Taylor \& Francis}
}

@article{rousseau2011asymptotic,
  title={Asymptotic behaviour of the posterior distribution in overfitted mixture models},
  author={Rousseau, Judith and Mengersen, Kerrie},
  journal={Journal of the Royal Statistical Society Series B: Statistical Methodology},
  volume={73},
  number={5},
  pages={689--710},
  year={2011},
  publisher={Oxford University Press}
}

@article{quenouille1949relation,
  title={A relation between the logarithmic, {Poisson}, and negative binomial series},
  author={Quenouille, Maurice H},
  journal={Biometrics},
  volume={5},
  number={2},
  pages={162--164},
  year={1949},
  publisher={JSTOR}
}

@article{chakraborty2023likelihood,
  title={Likelihood Ratio Test-Based Drug Safety Assessment using {R Package pvLRT}.},
  author={Chakraborty, Saptarshi and Markatou, Marianthi and Ball, Robert},
  journal={R Journal},
  volume={15},
  number={1},
  year={2023}
}

@article{akaike1974new,
  title={A new look at the statistical model identification},
  author={Akaike, Hirotugu},
  journal={IEEE Transactions on Automatic Control},
  volume={19},
  number={6},
  pages={716--723},
  year={1974},
  publisher={IEEE}
}

@article{schwarz1978estimating,
  title={Estimating the dimension of a model},
  author={Schwarz, Gideon},
  journal={The Annals of Statistics},
  pages={461--464},
  year={1978},
  publisher={JSTOR}
}

@article{ramkumar2016statin,
  title={Statin therapy: review of safety and potential side effects},
  author={Ramkumar, Satish and Raghunath, Ajay and Raghunath, Sudhakshini},
  journal={Acta Cardiologica Sinica},
  volume={32},
  number={6},
  pages={631},
  year={2016}
}

@article{gulani2017gadolinium,
  title={Gadolinium deposition in the brain: summary of evidence and recommendations},
  author={Gulani, Vikas and Calamante, Fernando and Shellock, Frank G and Kanal, Emanuel and Reeder, Scott B},
  journal={The Lancet Neurology},
  volume={16},
  number={7},
  pages={564--570},
  year={2017},
  publisher={Elsevier}
}

@article{ramalho2016gadolinium,
  title={Gadolinium toxicity and treatment},
  author={Ramalho, Joana and Ramalho, Miguel and Jay, Michael and Burke, Lauren M and Semelka, Richard C},
  journal={Magnetic Resonance Imaging},
  volume={34},
  number={10},
  pages={1394--1398},
  year={2016},
  publisher={Elsevier}
}

@misc{fda2017gbca,
  author       = {{U.S. Food and Drug Administration}},
  title        = {{FDA} Drug Safety Communication: {FDA} identifies no harmful effects to date with brain retention of gadolinium-based contrast agents for MRIs; review to continue},
  year         = {2017},
  note          = {https://www.fda.gov/drugs/drug-safety-and-availability/fda-drug-safety-communication-fda-identifies-no-harmful-effects-date-brain-retention-gadolinium}
}

@book{wood2017generalized,
  title={Generalized additive models: an introduction with R},
  author={Wood, Simon N},
  year = {2017},
  publisher={Chapman and Hall/CRC}
}






\end{document}